\documentclass[hidelinks]{article}
\usepackage{arxiv}
\usepackage{algorithm}
\usepackage{graphicx}
\usepackage{tcolorbox}
\usetikzlibrary{positioning, arrows.meta}
\usepackage{enumitem}
\usepackage{makecell}
\usepackage{hyperref}
\usepackage{subcaption}
\usepackage{algpseudocode}
\usepackage{color}
\definecolor{light-gray}{gray}{0.95}
\usepackage{rotating}
\usepackage{multirow}
\usepackage{amsmath}
\usepackage{pgfplots}
\pgfplotsset{compat=1.18}
\usepackage{tabularx}
\usepackage{booktabs}
\usepackage{amssymb}

\usepackage[maxbibnames=2]{biblatex} 
\AtBeginDocument{
  \providecommand\BibTeX{{%
    \normalfont B\kern-0.5em{\scshape i\kern-0.25em b}\kern-0.8em\TeX}}}

\begin{document}

\title{EXTree: Towards Supporting Explainability in Attribute-based Access Control}


\author{%
  Shanampudi Pranaya Chowdary  \\
  Indian Institute of Technology Kharagpur, India\\
  \texttt{pranayas@kgpian.iitkgp.ac.in}   \And
  Shamik Sural \\
  Indian Institute of Technology Kharagpur, India\\
  \texttt{shamik@cse.iitkgp.ac.in} 
}
\date{}
\renewcommand{\headeright}{}
\renewcommand{\undertitle}{}

\maketitle

\begin{abstract}
With increasing emphasis on transparency in digital governance, users expect more than silence when their access requests are denied by a system. However, authorization methods are notorious for their inability to provide any form of meaningful feedback under such situations. This paper shows a direction towards how the problem of explainability can be mitigated in the context of Attribute-based Access Control (ABAC), arguably the most researched topic in access control in recent years. We introduce \textit{EXTree}, which represents ABAC policies optimized for both fast evaluation (\underline{E}fficiency) and human-centric feedback (e\underline{X}plainability) in the form of a \underline{Tree}. Two strategic dimensions are investigated, namely, \emph{Feedback Evaluation Strategies} — how to craft actionable explanations when access is denied, and \emph{Tree Construction Strategies} — how the policy trees should be structured for efficient yet interpretable decisions. Through extensive experiments, we compare entropy-based, changeability-based, and randomly generated trees across multiple configurations. Our results demonstrate that EXTree, built for efficiency and interpretability, can bridge the gap between complex
authorization logic and human understanding.

\end{abstract}



\section{Introduction}
\label{sec:intro}
Access control research has evolved significantly over the years. While traditional models like Role-based Access Control (RBAC) \cite{rbac} work well with stable organizational hierarchies, modern computing environments require more flexible approaches. This led to the emergence of Attribute-based Access Control (ABAC) \cite{hu2014guide}, which evaluates access rights using fine-grained user attributes, resource properties, and environmental conditions. Practical implementations of ABAC, however, face the critical challenge of providing meaningful feedback when access is denied. In organizational settings, users frequently ask questions like "Why was I denied access?" and "What do I need to change?" \cite{kapadia2004know}. The need is for an automated, intelligent response to such queries without needing human intervention. 

The challenge lies not simply in explaining a denial, but in efficiently identifying feasible corrective actions that users can realistically perform. Na\"ively enumerating possible policy changes is computationally expensive and often produces impractical suggestions. This brings about the need for feedback mechanisms that balance actionability with system performance. Additionally, feedback should avoid revealing sensitive policy details.

Based on these challenges, there are three fundamental research questions in ABAC explainability as enumerated below:
\begin{enumerate}[label=(\roman*)]
    \item RQ1: What constitutes actionable feedback for denied ABAC requests
    \item RQ2: How to generate meaningful explanations that respect practical attribute constraints
    \item RQ3: How to provide feedback efficiently without compromising ABAC system performance
\end{enumerate}

To address these questions, we propose \textit{EXTree} (\underline{E}fficient and e\underline{X}plainable Tree), which  
is an explainability-aware architecture that integrates attribute changeability with efficient decision making.
The key innovation lies in organizing policies around attribute changeability costs, such that the system can identify feasible transitions as low-cost paths from \textit{deny} states to \textit{allow} states.
This approach facilitates structured, actionable feedback generation. We evaluate our framework using synthetic as well as realistic ABAC datasets, comparing various tree-construction strategies and feedback mechanisms. Our results demonstrate that change-aware tree construction significantly improves the quality of explanations without affecting computational efficiency, particularly in large-scale deployments where traditional approaches tend to become intractable.

The rest of the paper is organized as follows. In Section \ref{sec:prelim}, we introduce some of the relevant background material. The structure and design of EXTree is presented in Section \ref{sec:model}. Details of all the experiments and their results are discussed in Section \ref{sec:exprslt}. We review related work in Section \ref{sec:related} and finally conclude in Section \ref{sec:concl}. 

\section{Preliminaries}
\label{sec:prelim}
In this section, we introduce some of the background material relevant to our work. 

\subsection{ABAC Definition and Notations}
\label{subsec:prelims_abac}

The development of access control systems has evolved through several paradigms. Discretionary Access Control (DAC) introduced the concept of ownership-based permissions, where resource owners determine access rights \cite{sandhu1994access}. RBAC \cite{rbac} groups permissions into roles, improving scalability and reducing administrative overhead in organizational settings. As systems became more dynamic, Attribute-based Access Control (ABAC) emerged as a more flexible model \cite{hu2014guide}. Unlike the static role assignments used in RBAC, ABAC makes access decisions based on a rich set of attributes for users, resources, and the surrounding environment.
ABAC is comprised of the following fundamental components:
\begin{itemize}
    \item $U$: A finite set of users in the system
    \item $O$: A finite set of protected objects or resources
    \item $E$: A set of environmental conditions
    \item $OP$: The set of allowable operations on objects
\end{itemize}

Each entity category is characterized by its attribute sets:
\[
\begin{aligned}
UA &= \{a^u_1, a^u_2, \ldots, a^u_n\}, \quad
OA = \{a^o_1, a^o_2, \ldots, a^o_m\}, \\
EA &= \{a^e_1, a^e_2, \ldots, a^e_p\}
\end{aligned}
\]

For each attribute $a^x_i$ where $x \in \{u, o, e\}$, we define:
\begin{itemize}
    \item A value domain $V^x_i = \{v^x_{i1}, v^x_{i2}, \ldots, v^x_{ik}\}$
    \item A mapping function $f^x_i : X \rightarrow V^x_i \cup \{\#\}$ where $X \in \{U, O, E\}$
    \item The special value \# denotes undefined or unknown attributes
\end{itemize}

An access control policy $P = \{r_1, r_2, \ldots, r_l\}$ consists of rules, where each rule $r_i$ is formally represented as:
\[
r_i = \langle c^u_i, c^o_i, c^e_i, op_i \rangle
\]

Here, $c^x_i$ represents conjunctions of predicates over the respective attribute sets, and $op_i \in OP$ specifies the permitted operation. Each predicate is of the form $a^x_j \ \text{rel} \ v$, where $\text{rel} \in \{ =, \neq, <, >, \le, \ge \}$ and $v \in V^x_j \cup \{\#,*\}$, with ``$*$'' denoting a wildcard that matches any value. An access request $q$ is a mapping from a subset of attributes in $UA \cup OA \cup EA$ to concrete values in their domains. Evaluation of $q$ against $P$ determines which rule(s) $r_i \in P$ are satisfied.

To ground the above notation, consider a minimal ABAC policy with user attributes
\[
UA = \{\text{role}, \text{department}, \text{clearance}, \text{training\_over}\}.
\]

For this policy, access is permitted under the following rules:
\begin{itemize}
    \item $r_1 = \langle \text{role} = \text{admin}, *, *, op \rangle$
    \item $r_2 = \langle \text{role} = \text{intern} \land \texttt{clearance} = \text{medium}, *, *, op \rangle$
    \item $r_3 = \langle \text{role} = \text{manager} \land \text{clearance} = \text{low} \land \text{department} = \text{HR}, *, *, op \rangle$
\end{itemize}

Consider the access request
$q = (\text{role} = \text{manager} \land \text{clearance} = \text{medium} \land \text{department} = \text{HR}, *, op$),
which is denied under this policy.

\subsection{Policy Evaluation Architectures}
\label{subsec:eval_architectures}

The industry standard XACML provides a widely used framework for implementing ABAC policies~\cite{xacml_spec}. It defines core components such as the Policy Decision Point (PDP) and Policy Enforcement Point (PEP), which together handle policy evaluation and enforcement. In practice, however, the performance of the PDP can become a bottleneck as the number and complexity of policy rules increase~\cite{butler_xacml}.

Several formal approaches have been explored to improve policy analysis and verification. For example, Ordered Binary Decision Diagrams (OBDDs) provide compact and canonical representations that support efficient policy analysis~\cite{margrave}. Answer Set Programming (ASP) has also been used to reason about properties such as policy completeness and conflict detection~\cite{asp_xacml}. While these approaches improve correctness and verifiability, they often introduce additional computational overhead.

The PolTree framework~\cite{nath2019poltree} takes a different approach by organizing policies in a hierarchical structure. Instead of evaluating policies sequentially, PolTree represents them as a decision tree where internal nodes correspond to attributes and leaf nodes represent access decisions. This structure allows requests to be evaluated more efficiently while also making it easier to trace how a particular decision was reached.
Experimental results demonstrate significant performance improvements over conventional methods, particularly in systems with a large number of rules. 



\section{Structure and Design of EXTree}
\label{sec:model}
We argue that a sequential representation and enforcement of ABAC rules as described in Section \ref{subsec:prelims_abac}, are not amenable to efficient evaluation or towards meaningful and graded explainability. Hence, hierarchical structuring of rules is imperative so that some form of ordering can be imposed on the list of attributes. Towards this, we propose EXTree that attempts to capture the required features in a unified manner. This section describes the structure of EXTree and how it is built, along with the relevant design considerations. It also presents the algorithm for supporting explainability in EXTree.

\subsection{EXTree Representation}
\label{subsec:extree_representation}

\begin{figure}[t]
    \centering
    \includegraphics[width=0.9\linewidth]{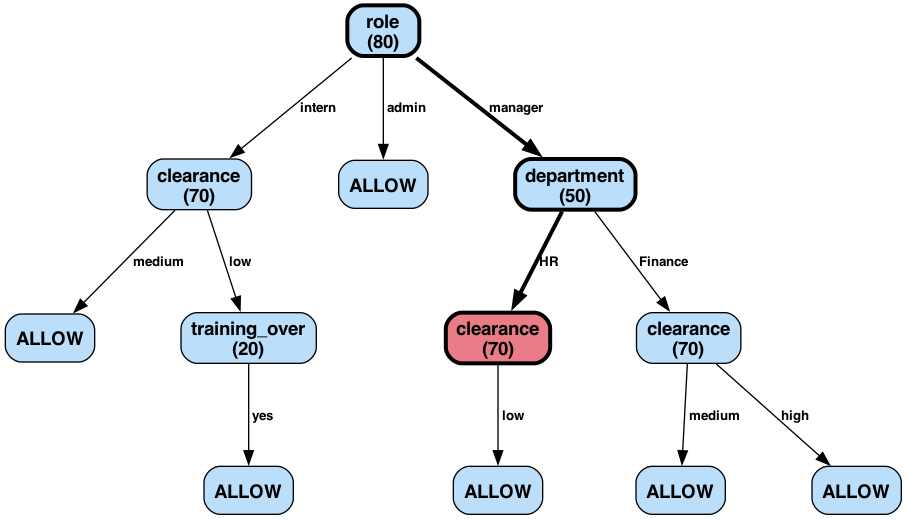}
    \caption{\footnotesize Hierarchical representation of the example ABAC policy introduced in Section~\ref{subsec:prelims_abac}. Internal nodes correspond to attribute tests and leaf nodes represent \textit{allow} decisions. The highlighted (in bold) path illustrates the evaluation trace for the denied request and, the deny node is marked in red.}
    \label{fig:poltree_example}
\end{figure}

An EXTree $T$ is an $n$-ary hierarchical structure for evaluation of ABAC policies as shown in Figure \ref{fig:poltree_example}. 
This structural organization has some similarity with the PolTree framework~\cite{nath2019poltree}, which also arranges ABAC policies as decision trees over attributes for efficient evaluation. 
EXTree retains this core representation but extends it with explainability-oriented metadata and traversal mechanisms, as described in the following sections.
The structure of EXTree satisfies the following criteria.

\begin{itemize}
    \item Each non-leaf node is labeled with an attribute $a \in UA \cup OA \cup EA$.
    \item Each outgoing edge from a node represents a possible value or predicate outcome for that attribute.
    \item Each leaf node stores the subset of policy rules from $P$ consistent with the predicates along its root–leaf path.
\end{itemize}

Evaluation of an access request $q$ proceeds top–down 
starting from the root. Each node’s attribute is compared against the corresponding value in $q$ with traversal continuing along the matching edge. If no matching edge or wildcard exists, the search halts and the decision defaults to \textit{deny}, reflecting that no rule in $P$ satisfies all constraints along the path.
\begin{equation}
\text{Decision}(q) =
\begin{cases}
\text{allow}(op), & \text{if a path to an \textit{allow} leaf exists;}\\
\text{deny}, & \text{otherwise.}
\end{cases}
\end{equation}
As the system models only \textit{allow} policies, each leaf corresponds to one or more permissible operations. When an access request $q$ reaches a leaf, the operations stored there define the actions granted to the requester. Denial arises implicitly from the absence of a valid path from the root to a leaf level node.

If wildcard branches are present, evaluation may backtrack when the exact-value path fails, exploring more general branches.  
This \textit{retracing} mechanism ensures that specific rules take precedence, while generalized ones apply only when no exact match is found. 
Retracing introduces negligible overhead since wildcard branches are rare, keeping search time approximately linear in the number of attributes. 

At each internal node, exactly one attribute is tested, and outgoing branches correspond to mutually exclusive values of that attribute (apart from optional wildcards). Each rule is placed only in subtrees consistent with its predicates. Therefore, a request that satisfies a rule will necessarily traverse a path that allows that rule.

Figure~\ref{fig:poltree_example} shows the hierarchical representation of the policy introduced in Section~\ref{subsec:prelims_abac}. Internal nodes correspond to attribute tests, while leaf nodes represent \textit{allow} decisions. Requests are evaluated by traversing the tree top--down until a matching leaf is reached or no valid path exists. For simplicity, we assume a single operation for all rules and requests in this example and hence, the operation attribute is omitted from this EXTree representation.

To enable explainability and actionable feedback (i.e., to address research question RQ1), we introduce a meta-policy $M$ in EXTree comprised of two functions~\cite{cue_framework} that augment the base ABAC model:

\subsubsection*{i. Visibility Function}
\begin{equation}
\text{Vis}(a, \text{actor}) \in [0,1]
\end{equation}
quantifies how much information about attribute $a$ may be disclosed to a particular actor (user, administrator, or requester).  
A value of 1 indicates full visibility, 0 indicates full confidentiality, and intermediate values encode partial or probabilistic disclosure.

\subsubsection*{ii. Change Cost Function}
\begin{equation}
\text{ChangeCost}(a, v_{\text{from}}, v_{\text{to}}) \in [0, \infty)
\end{equation}
represents the estimated difficulty or effort required to modify an attribute $a$ from $v_{\text{from}}$ to $v_{\text{to}}$.  
This may be accounting for monetary cost, time or administrative work required.  
When generating feedback, the system focuses on attributes that the requester can realistically modify. Attributes with lower change costs are therefore preferred, subject to the constraints defined by the visibility function earlier.

A normalized attribute-level measure can be defined as:
\begin{equation}
C(a) = \frac{c_a - c_{\min}}{c_{\max} - c_{\min}}, \qquad
\text{Changeability}(a) = 100(1 - C(a))
\end{equation}
where $c_a$ is the raw assigned difficulty score and $c_{\min}, c_{\max}$ are system-specific bounds.

\begin{table}[t]
\centering
\footnotesize
\caption{Default Change Cost Profiles}
\label{tab:defaultcosts}
\renewcommand{\arraystretch}{1.4
}
\begin{tabular}{llc}
\hline
\textbf{Category} & \textbf{Characteristics} & \textbf{Cost Range} \\
\hline
Environmental (E) & High volatility, system-controlled & 0-40 \\
User (U) & Moderate mutability, administrative oversight & 50-90 \\
Object (O) & Low mutability, security-critical & 80-100 \\
\hline
\end{tabular}
\end{table}

Table ~\ref{tab:defaultcosts} lays down a fallback set of values for attributes if required. This baseline reflects a pragmatic assumption: environmental conditions are transient and system-controlled, user attributes are modifiable but bureaucratic, while object attributes are largely immutable once provisioned.  
Such default priors can later be adapted via monitoring, feedback or empirical policy usage statistics.

\subsection{EXTree Design Considerations}
\label{subsec:EXTreeDesignConsiderations}

In this sub-section, we present two major design considerations while building an EXTree from ABAC rules.

\subsubsection{Attribute Splitting Criterion}
\label{subsubsec:splittingcriteria}
Since EXTree inherits its core hierarchical representation from the PolTree framework~\cite{nath2019poltree}, a natural baseline for tree construction is the entropy-based splitting heuristic commonly used in decision tree.
Entropy measures the uncertainty or impurity of a dataset—zero when all samples share the same outcome, and a maximum value when classes are evenly split. In decision trees, it quantifies how mixed the decision outcomes are at a node:
\begin{equation}
H(S) = -p_{\text{allow}}\log_2(p_{\text{allow}}) - p_{\text{deny}}\log_2(p_{\text{deny}}), \quad H(S)\in[0,1]
\end{equation}
\noindent
where $p_{\text{allow}}$ and $p_{\text{deny}}$ denote the proportions of allowed and denied requests. The splitting attribute is typically chosen to maximize information gain—the largest reduction in entropy:
\begin{equation}
\text{Gain}(S, A) = H(S) - \sum_{v \in \text{Values}(A)} \frac{|S_v|}{|S|} H(S_v)
\end{equation}
\noindent
Each internal node represents an attribute predicate (e.g., \texttt{Departme-\\nt = HR}), and the leaves denote \texttt{permit}/\texttt{deny} outcomes. The \textit{Highest-Entropy-First} heuristic seeks attributes that most evenly divide policies, aiming for shallower and faster evaluation trees.

However, ABAC attributes are often semantically correlated (for example, \texttt{role} and \texttt{clearance}), so the entropy across attributes do not always differ much. Moreover, in EXTree each level evaluates a distinct attribute, which means the depth of the tree is naturally limited by the number of attributes available. Because of these properties, we hypothesize that entropy-based splitting may not offer much advantage beyond what the hierarchical structure itself achieves. To explore this further, we compare several alternative splitting strategies: \textit{Highest-Entropy-First}, \textit{Lowest-Entropy-First}, \textit{Changeability-Based}, and \textit{Random}.


\subsubsection{Feedback Strategy Evaluation}
\label{subsubsec:feedback-strategy}

\begin{figure*}[t]
    \centering
    \begin{subfigure}{0.45\textwidth}
    \includegraphics[width=\linewidth]{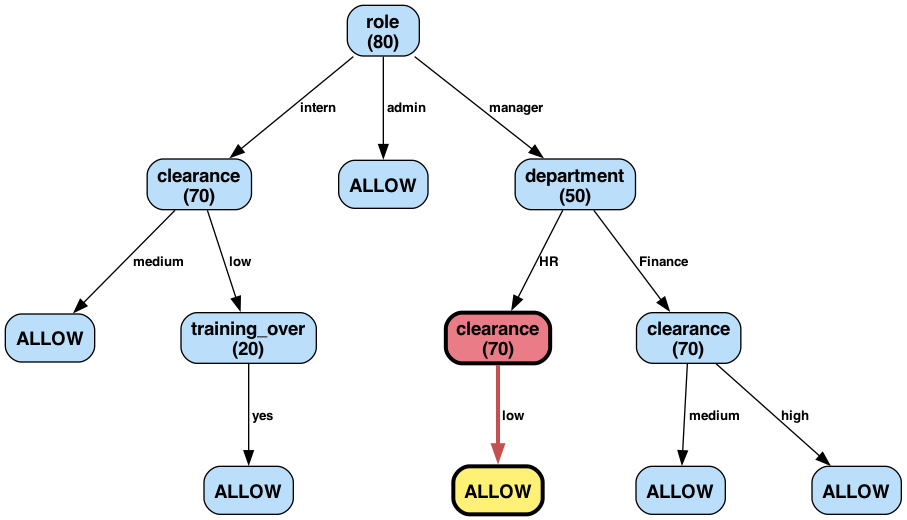}
    \caption{\footnotesize Depth-first}
     \label{fig:depth-first}
    \end{subfigure}
    \begin{subfigure}{0.45\textwidth}
        \includegraphics[width=\linewidth]{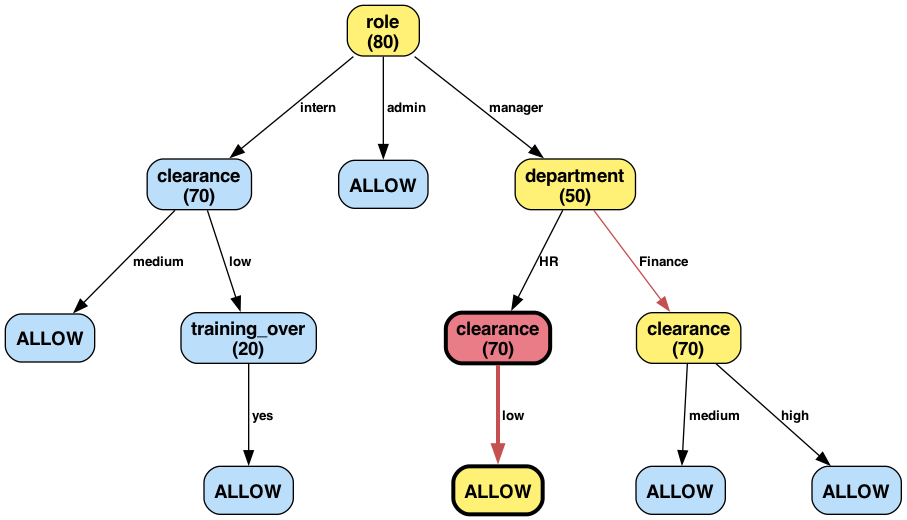}
        \caption{\footnotesize Depth-best}
        \label{fig:depth-best}
    \end{subfigure}
    \begin{subfigure}{0.45\textwidth}    
    \includegraphics[width=\linewidth]{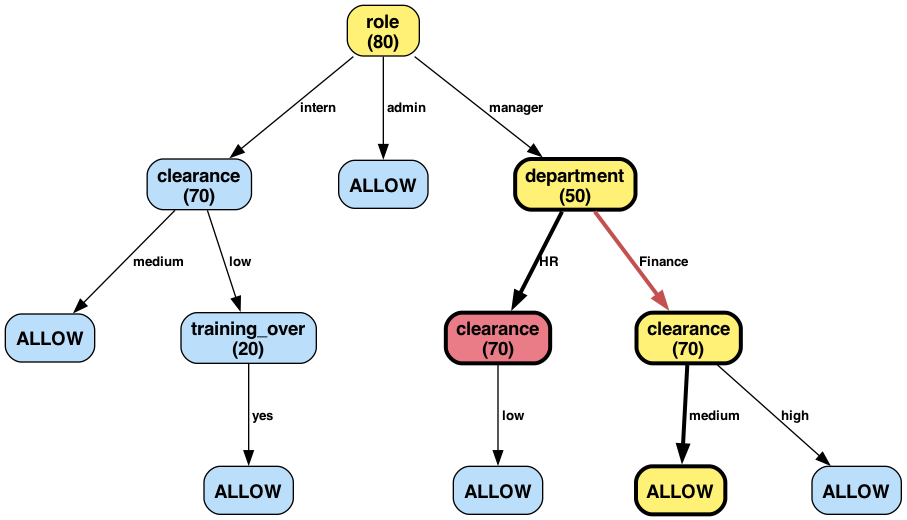}
    \caption{\footnotesize Change-first}
    \label{fig:change-first}
    \end{subfigure}
    \begin{subfigure}{0.45\textwidth}
    \includegraphics[width=\linewidth]{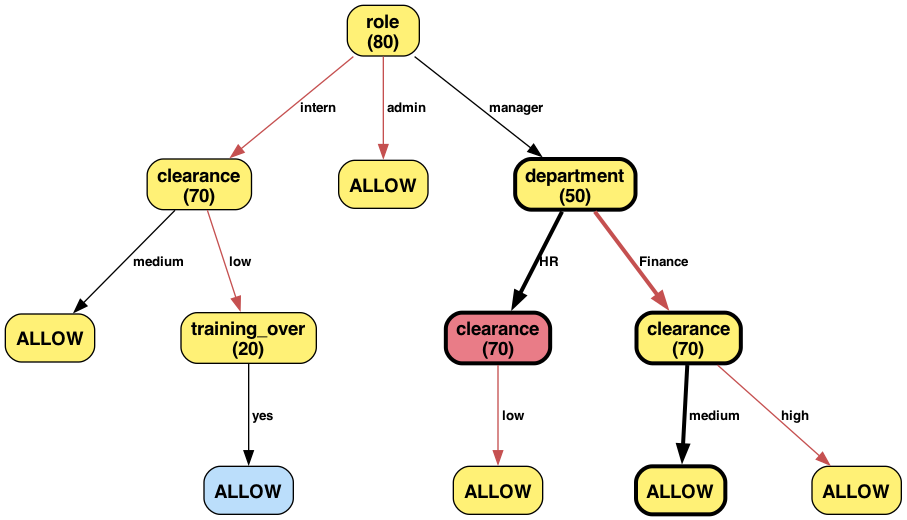}
    \caption{\footnotesize Change-best}
    \label{fig:change-best}
\end{subfigure}
    
    \caption{Visualization of feedback search strategies on the example EXTree. The denied request (\text{role = manager}, \text{department = HR}, \text{clearance = medium}) terminates at the deny \textit{red} node. Visited nodes are highlighted in yellow, red edges denote attribute deviations from the request, and the selected feedback path is shown in bold. The feedback cost is the sum of the changeability costs along the selected path.}
    \label{fig:traversal examples}
   
\end{figure*}

To address RQ2, this sub-section introduces four feedback generation strategies for EXTree that explore different trade-offs between \textit{feedback interpretability} with \textit{computational efficiency}. Each strategy begins at a \textit{deny node} and searches the EXTree for reachable \textit{allow} nodes through permissible attribute modifications (Figure~\ref{fig:poltree_example}). The deny node captures the deepest point in the policy tree that the request satisfies, making it the natural starting point for generating minimal corrective feedback.

\paragraph{Actionable Feedback.}
In this work, we define \emph{actionable feedback} for a denied ABAC request as a bounded set of attribute modifications that transforms the request from a deny outcome to an allow outcome. Among feasible explanations, lower cumulative changeability cost corresponds to more actionable feedback, as it prioritizes attributes that users can realistically modify. In practice, attributes vary in mutability due to organizational constraints: attributes such as roles or security clearances are often fixed, whereas contextual or procedural attributes may be modified with lower effort. Changeability cost provides a principled way to distinguish feasible modifications from impractical ones. Accordingly, EXTree is primarily designed to support near-miss denials—cases where a request narrowly fails policy constraints and feasible corrective modification is realistic. While the primary focus is user-facing remediation, the EXTree structure and feedback strategies also support administrative analysis and policy debugging.

Two search-limiting constraints ensure tractability for high-dimensional ABAC policies:
\begin{itemize}
    \item \textbf{Depth limit} (\textit{max\_depth}): restricts traversal distance from the deny node
    \item \textbf{Change limit} (\textit{max\_changes}): restricts maximum attribute modifications in feedback
\end{itemize}

Depending on the termination policy, strategies either return the first valid \textit{allow} node encountered (\emph{greedy}) or exhaustively evaluate all feasible candidates and return the minimum-cost solution. The four strategies evaluated are \textbf{depth-first}, \textbf{depth-best}, \textbf{change-first}, \textbf{change-best}.

\paragraph{Algorithm and Complexity}
\label{subsec:feedback_search}

Algorithm~\ref{alg:feedback_search} performs a bounded search around the deny node to identify reachable \textit{allow} nodes that require minimal attribute modifications. The EXTree is treated as a graph, allowing traversal both upward (constraint relaxation) and downward (constraint addition). Change cost is accumulated only when traversing to child nodes whose predicates are violated by the request; traversal toward parent nodes does not increase cost.

The search is constrained by a maximum traversal depth $D=\texttt{max\_depth}$ and a maximum number of attribute changes $K=\texttt{max\_changes}$. The choice of strategy determines the candidate expansion order, prioritizing either proximity to the deny node (depth-based) or lower cumulative changeability cost (change-based). Greedy variants terminate upon encountering the first valid \textit{allow} node, while exhaustive variants enumerate all feasible candidates within these bounds.

Let $A$ denote the number of distinct attributes in the policy and $b$ the maximum branching factor of the EXTree. Since each level of the tree corresponds to a test on a distinct attribute, the tree height is bounded by $A$. In the worst case, a depth-limited traversal explores $O(b^D)$ nodes. Independently, the number of feasible change configurations is bounded by the number of ways to select and modify up to $K$ attributes out of $A$, where each modified attribute can take up to $b$ values:
\[
\sum_{k=0}^{K} \binom{A}{k} b^k = O\!\left((A b)^K\right).
\]
Accounting for traversal across at most $A$ levels of the tree for each change configuration, the total number of explored states is bounded by
\[
M = O\!\left(\min(b^D,\; A^{K+1} b^K)\right).
\]

Change-based strategies incur an additional $O(\log M)$ overhead due to priority queue operations, whereas depth-based strategies rely on constant-time stack operations. Although the worst-case search is exponential in bounds $D$ and $K$, in practice, early termination and tight change limits significantly reduce the explored search space, as observed empirically. Particularly since typical branching factors are modest ($b \approx 2$-$5$) and change limits are small ($K \leq 3$).

\algtext*{EndIf}
\algtext*{EndFor}
\algtext*{EndWhile}

\begin{algorithm}
\caption{Feedback Path Search in EXTree}
\label{alg:feedback_search}
\begin{algorithmic}[1]
\State \textbf{Input:} Policy tree $T$, meta-policy, deny node $n_d$, request $r$, user $u$, strategy $\mathcal{S}$, limits $(\texttt{max\_depth}, \texttt{max\_changes})$
\State \textbf{Output:} $(S^{*}, c^{*})$ where $S^{*}$ is the suggested attribute changes and $c^{*}$ is the total cost, or \texttt{null}

\State Initialize candidate collection $\mathcal{C}$:
\State \hspace{1em} stack if $\mathcal{S}\in\{\texttt{depth\_first},\texttt{depth\_best}\}$,
priority queue (by increasing cost) if $\mathcal{S}\in\{\texttt{change\_first},\texttt{change\_best}\}$
\State Insert $(n_d, 0, 0, \emptyset)$ into $\mathcal{C}$ \Comment{(node, cost, depth, change set)}
\State $c^{*} \leftarrow \infty,\ S^{*} \leftarrow \emptyset$

\While{$\mathcal{C}$ not empty}
    \State Pop $(n, c, d, S)$ from $\mathcal{C}$ \Comment{best according to strategy}

    \If{$n$ visited}
        \State \textbf{continue}
    \EndIf
    \State Mark $n$ visited

    \If{$d > \texttt{max\_depth}$ or $|S| > \texttt{max\_changes}$}
        \State \textbf{continue}
    \EndIf

    \If{$n$ is \textbf{ALLOW}}
        \If{$\mathcal{S}\in\{\texttt{depth\_first},\texttt{change\_first}\}$}
            \State \Return $(S, c)$
        \ElsIf{$c < c^{*}$}
            \State $c^{*} \leftarrow c,\ S^{*} \leftarrow S$
        \EndIf
    \EndIf

    \If{$n$ has a parent $p$}
        \Comment{Relax constraint: move to parent (no cost added)}
        \State Insert $(p, c, d+1, S)$ into $\mathcal{C}$
    \EndIf

    \State Let $a$ be the attribute tested on edge $(n \rightarrow \cdot)$
    \Comment{Add constraint: move to children (may incur cost)}
    \ForAll{children $n'$ of $n$}
        \If{visibility[a][u] = 0}
            \State \textbf{continue}
        \EndIf
        
        \State $c' \leftarrow c,\ S' \leftarrow S$
        \If{$r[a]$ does not equal the value on edge $(n \rightarrow n')$}
            \State $c' \leftarrow c + \text{change\_cost}[a]$
            \State $S' \leftarrow S \cup \{a \colon r[a]\}$
        \EndIf
        \State Insert $(n', c', d+1, S')$ into $\mathcal{C}$
    \EndFor
\EndWhile

\State \Return $(S^{*}, c^{*})$
\end{algorithmic}
\end{algorithm}

\subsubsection*{Feedback Strategies}
\label{subsubsc:feedback_strategies}

The four feedback strategies represent distinct philosophies of search control—ranging from greedy, time-sensitive heuristics to exhaustive, globally optimal traversals. We illustrate their behavior using the same policy and denied request introduced earlier. Figure~\ref{fig:poltree_example} shows the corresponding EXTree and the evaluation trace. We assign fixed changeability costs to attributes to reflect their relative difficulty of modification. Specifically, the costs used in the example are: \text{training\_over} = 20, \text{department} = 50, \text{clearance} = 70, and \text{role} = 80. The search is constrained by \textit{max\_changes} = 2 and \textit{max\_depth} = 3. 


    
   

\subsubsection*{Depth-first Search}
\label{subsubsec:depthfirst}

This strategy traverses the EXTree in a depth-first manner and terminates immediately upon encountering the first \textit{allow node} within the allowed depth limit. It prioritizes speed over optimality, generating fast but potentially suboptimal feedback as shown in Figure \ref{fig:depth-first}.
For the example request, depth-first search suggests modifying \texttt{clearance} from \texttt{medium} to \texttt{low}, producing an \textit{allow} decision with a cumulative changeability cost of $70$.

\subsubsection*{Depth-best Search}
\label{subsubsec:depthbest}

Unlike the greedy variant, the depth-best strategy performs a complete traversal of all candidate paths within the given depth limit and returns the \textit{allow node} with the minimal cumulative changeability cost  as shown in Figure \ref{fig:depth-best}. It provides more accurate and interpretable feedback at the expense of higher computational overhead.
For the example request, depth-best search also suggests \texttt{clearance} from \texttt{medium} to \texttt{low}, producing an \textit{allow} decision with a cumulative changeability cost of $70$.


\subsubsection*{Change-first Search}
\label{subsubsec:changefirst}

This strategy prioritizes attribute modification feasibility over structural traversal depth. Starting from the deny node, it explores changes to low-cost attributes first and terminates upon reaching the first valid \textit{allow node} that respects the \textit{max\_changes} constraint as shown in Figure \ref{fig:change-first}. 
For the example request, change-first search suggests changing the \texttt{department} attribute from \texttt{HR} to \texttt{Finance}, resulting in an \textit{allow} decision with a cumulative changeability cost of $50$.


\subsubsection*{Change-best Search}
\label{subsubsec:changebest}
The exhaustive variant searches all reachable \textit{allow nodes} within the change limit and returns the minimum-cost modification as shown in Figure \ref{fig:change-best}. This provides globally optimal feedback and is suitable for administrative or audit settings where completeness matters more than response time. For the example request, this exhaustive search also identifies modifying \texttt{department} from \texttt{HR} to \texttt{Finance} as the optimal correction, resulting in a cost of $50$.

As is evident from the design of EXTree, RQ1 is sufficiently addressed by formally defining what constitutes actionable feedback in the ABAC setting.

\paragraph{Visibility-Constrained Feedback.}

In realistic organizational settings, certain policy rules encode sensitive information—internal classification thresholds, administrative roles, or confidential resource attributes—that cannot be disclosed without violating security constraints.
Visibility constraints are inherently user-dependent, whereas EXTree construction is policy-centric and user-agnostic. Incorporating visibility during tree construction would therefore require rebuilding the tree per user, which is impractical. We instead enforce visibility as an evaluation-time constraint during feedback generation. Our current implementation uses a binary visibility model, where attributes are either visible or hidden. This captures the common use case, while richer visibility models are left for future work.

Each edge corresponding to an attribute–value predicate $a$ is assigned a binary visibility flag $v_a \in \{0,1\}$. For a candidate explanation path $P = \{a_1,\dots,a_k\}$, the visibility-aware cost is:
\[
C_v(P)=
\begin{cases}
\infty, & \exists a_i \in P \text{ s.t. } v_{a_i}=0,\\
\sum_{i=1}^k C(a_i), & \text{otherwise}.
\end{cases}
\]
Paths traversing hidden predicates are excluded from consideration by assigning infinite traversal cost during evaluation, without modifying the underlying EXTree structure.

\section{Experimental Results}
\label{sec:exprslt}

In this section, we carry out extensive experiments to study the performance of EXTree under various conditions.

\subsection{Datasets}
\label{subsec:datasets}

We evaluate EXTree using two synthetic ABAC datasets and one realistic dataset from the ABACLab repository~\cite{abaclab_github}.

\paragraph{Synthetic datasets.}  
Two datasets were generated using the publicly available generator of~\cite{paul2021efficient}, augmented with 30--50\% wildcard predicates to simulate realistic policy overlap. Synthetic--1 provides a balanced baseline (1000 users, 1000 objects, 1000 policies), while Synthetic--2 scales entity and policy counts to evaluate performance under larger rule sets. Both maintain an approximate 70:30 allow-to-deny ratio.

\paragraph{Realistic dataset.}  
The Healthcare Access dataset from ABACLab~\cite{abaclab_github,bui2025abaclab} models fine-grained access control in a healthcare setting, incorporating role hierarchies, sensitivity levels, and contextual attributes. Although smaller in size, it captures realistic attribute correlations and policy semantics.

Further details regarding parameter choices and generator setup are provided in Appendix~\ref{dataset_details}.

\subsection{Effect of Splitting Criterion on EXTree Efficiency}
\label{subsec:splittingcriterion}
This section studies the effect of different splitting criteria on the efficiency of EXTree. Each dataset was evaluated using 10,000 random queries simulating realistic access requests. Metrics recorded include tree generation time, total number of nodes, average decision latency, and total query time.

In Tables \ref{tab:synthetic-results} and \ref{tab:abac-results}, we report results for the splitting heuristics introduced in Section \ref{subsubsec:splittingcriteria}. Across datasets and metrics, no heuristic consistently outperforms the others. While \textit{Lowest-Entropy-First} trees often contain fewer nodes, this does not translate to lower query latency. In some cases, smaller trees give rise to higher decision times due to unbalanced partitions, whereas larger trees benefit from more uniform branching.

Overall, the performance of all heuristics falls within a narrow range. This suggests that the main efficiency gain comes from the hierarchical organization itself rather than the specific splitting rule. Such a behavior can be explained by the structural properties of ABAC policy trees. Since each level tests a distinct attribute, the tree depth is bounded by the total number of attributes. This ensures that different splitting heuristics cannot create very deep or highly skewed trees, limiting the benefits of entropy-based optimization.

In addition, many ABAC attributes are semantically related (e.g., department, clearance, resource type), which limits variation across possible splits.

\setlength{\tabcolsep}{3.5pt}
\begin{table}[h!]
\centering
\footnotesize
\caption{Synthetic Dataset Performance Comparison for Four Splitting Criteria}
\renewcommand{\arraystretch}{1.4}
\begin{tabular}{lcccccc}
\toprule
\textbf{Dataset} & \textbf{Heuristic} & \textbf{Nodes} & \textbf{Allowed} & \textbf{Denied} & \textbf{Avg. Time (s)} \\
\midrule
\multirow{4}{*}{Synthetic-1} 
& Highest Entropy & 13,983 & 6,976 & 3,024 & $\boldsymbol{1.68\times10^{-5}}$ \\
& Lowest Entropy & \textbf{11,999} & 6,976 & 3,024 & $2.93\times10^{-5}$ \\
& Highest Change Cost & 14,216 & 6,976 & 3,024 & $2.93\times10^{-5}$\\
& Random & 14,205 & 6,976 & 3,024 & $2.93\times10^{-5}$\\ \midrule

\multirow{4}{*}{Synthetic-2} 
& Highest Entropy & 12,090 & 7,037 & 2,963 & $3.23\times10^{-5}$ \\
& Lowest Entropy & \textbf{9,753} & 7,037 & 2,963 & $3.64\times10^{-5}$\\
& Highest Change Cost & 14,187 & 7,037 & 2,963 & $2.89\times10^{-5}$ \\
& Random & 14,208 & 7,037 & 2,963 & $\boldsymbol{1.72\times10^{-5}}$  \\ 
\bottomrule
\end{tabular}
\label{tab:synthetic-results}
\end{table}

\setlength{\tabcolsep}{2.5pt}
\begin{table}[h!]
\centering
\footnotesize
\caption{Realistic ABAC Dataset Performance Comparison for Four Splitting Criteria}
\renewcommand{\arraystretch}{1.4}
\begin{tabular}{lcccccc}
\toprule
\textbf{Heuristic} & \textbf{Nodes} & \textbf{Allowed} & \textbf{Denied} & \textbf{Avg. Time (s)} & \textbf{Total Time (s)} \\
\midrule
Lowest Entropy & \textbf{250} & 41 & 34 & $\boldsymbol{3.45\times10^{-6}}$ & $\boldsymbol{2.58\times10^{-4}}$ \\
Highest Entropy & 361 & 41 & 34 & $4.78\times10^{-6}$ & $3.58\times10^{-4}$ \\
Highest Change Cost & 254 & 41 & 34 & $5.56\times10^{-6}$ & $4.17\times10^{-4}$ \\
Random & 320 & 41 & 34 & $4.74\times10^{-6}$ & $3.56\times10^{-4}$ \\
\bottomrule
\end{tabular}
\label{tab:abac-results}
\end{table}

We see that the choice of splitting criterion has negligible impact on runtime performance. This allows EXTree construction to prioritize interpretability metrics such as \textit{changeability cost} or \textit{attribute visibility}—without sacrificing decision-making efficiency. The main improvement in access evaluation time comes from the hierarchical structure itself, allowing EXTree to optimize for interpretability without affecting performance.

\subsection{Feedback Strategy Results}
\label{subsec:feedbackstrategyresults}

The feedback generation strategies depend on two parameters: the maximum traversal depth (\texttt{max\_depth}) and the maximum number of allowed attribute modifications (\texttt{max\_changes}). These parameters govern the trade-off between feedback coverage, complexity, and computational cost, and must be selected on a per-dataset basis due to variations in policy size and structure. The method used for selecting these parameters, and their overall effect on results is given in Appendix~\ref{subsubsec:paramopt}.

\subsubsection*{Strategy Evaluation}
\label{subsubsec:strategyevaluation} 
Using the identified parameters,
we now evaluate and compare the four feedback generation strategies introduced in Section \ref{subsubsec:feedback-strategy}. The evaluation uses the following six metrics:

\begin{itemize}
    \item \textbf{Average Cost:} Average cumulative changeability cost of modified attributes. Lower costs indicate more feasible changes.
    \item \textbf{Average Time (ms):} Average time per denied request, measuring computational efficiency.
    \item \textbf{Found Fraction:} Ratio of denied requests yielding valid feedback paths. A value of 1.0 indicates complete coverage.
    \item \textbf{Nodes Expanded:} Average number of nodes visited during traversal, measuring search complexity.
    \item \textbf{Average Depth:} Average distance between \textit{deny} and the final \textit{allow} node chosen.
    \item \textbf{Average Number of Changes:} Average number of attributes needing modification. Lower values mean more actionable feedback.
\end{itemize}


Experiments were performed on all datasets using two tree types: \textit{entropy-based trees} (optimized for decision efficiency) and \textit{high-cost-first trees} (optimized for interpretability).
Tables~\ref{tab:synthetic_8_30} and ~\ref{tab:healthcare_abac} present detailed performance statistics. 

The Synthetic-1 dataset is omitted from the tabulated results. For this dataset, entropy-based trees fail to produce actionable feedback under the selected traversal and modification limits, resulting in a zero success rate across all strategies. 

Notably, change-aware tree constructions on the same dataset still achieve high feedback coverage (0.95). This highlights the role of the policy tree structure in supporting explainability, and addresses RQ2 on how meaningful explanations can be generated under practical attribute constraints.

Two consistent trends emerge across all datasets:

\begin{enumerate}[label=(\roman*)]
    \item \textbf{Superiority of Change-first:} The \texttt{change-first} strategy consistently achieves the lowest computation time without sacrificing explanation quality. Because the search proceeds greedily over attributes, feasible low-cost corrections are typically found early. In deeper trees, it avoids unnecessary exploration, and typically expands fewer nodes than alternative strategies.

    \item \textbf{Synergy with High-Cost-First Trees:} When the tree is structured using high change-cost attributes near the root, \texttt{change\\-first} becomes even more effective. The attribute ordering guides the search toward simpler corrections, often resulting in shallow traversal depth and only a few attribute modifications. In practice, this produces concise and interpretable feedback.
\end{enumerate}

Depth-based strategies occasionally achieve slightly better optimal costs, but they do so by expanding many more nodes, which introduces significant computational overhead. In contrast, \texttt{change\\-first} provides a better balance between efficiency and explanation quality, making it more suitable for real-time decision support. The strong coverage and low-cost feedback produced by change-based strategies show how meaningful explanations can be generated under practical constraints, addressing research questions RQ1 and RQ2.

\begin{table*}[t]
\centering
\footnotesize
\setlength{\tabcolsep}{4pt} 
\renewcommand{\arraystretch}{1.15}
\caption{Performance of different feedback strategies on the Synthetic-2 dataset ($\texttt{max\_depth}=30$, $\texttt{max\_changes}=8$).}
\label{tab:synthetic_8_30}
\begin{tabular}{l l r r r r r r}
\toprule
\textbf{Tree Type} & \textbf{Strategy} &
\textbf{Avg. Cost} & \textbf{Avg. Time (ms)} &
\textbf{Found Fraction} & \textbf{Nodes Expanded} &
\textbf{Avg. Depth} & \textbf{Avg. \#Changes} \\
\midrule
\multirow{4}{*}{Entropy}
 & Depth-first  & 490.16 & \textbf{17.47} & \textbf{0.82} & \textbf{5198.95} & \textbf{15.99} & 7.62 \\
 & Depth-best   & \textbf{462.68} & 22.21 & \textbf{0.82} & 6789.90 & 18.00 & \textbf{7.49} \\
 & Change-first & \textbf{462.68} & 25.93 & \textbf{0.82} & 5858.08 & 18.00 &\textbf{ 7.49} \\
 & Change-best  & \textbf{462.68} & 29.00 & \textbf{0.82} & 6789.92 & 18.00 & \textbf{7.49} \\
\midrule
\multirow{4}{*}{High-Cost-First}
 & Depth-first  & 322.51 & \textbf{1.98}  & \textbf{1.00} & 520.68  & \textbf{8.54}  & 6.62 \\
 & Depth-best   & \textbf{213.63} & 22.89 & \textbf{1.00} & 6960.61 & 14.05 & \textbf{4.02} \\
 & Change-first & \textbf{213.63} & 2.42  & \textbf{1.00} & \textbf{460.75}  & 14.05 & \textbf{4.02} \\
 & Change-best  & \textbf{213.63} & 29.60 & \textbf{1.00} & 6960.61 & 14.05 & \textbf{4.02} \\
\bottomrule
\end{tabular}
\end{table*}

\begin{table*}[t]
\centering
\footnotesize
\setlength{\tabcolsep}{4pt} 
\renewcommand{\arraystretch}{1.15}
\caption{Performance of different feedback strategies on the ABACLab Healthcare dataset ($\texttt{max\_depth}=5$, $\texttt{max\_changes}=3$).}
\label{tab:healthcare_abac}
\begin{tabular}{l l r r r r r r}
\toprule
\textbf{Tree Type} & \textbf{Strategy} &
\textbf{Avg. Cost} & \textbf{Avg. Time (ms)} &
\textbf{Found Fraction} & \textbf{Nodes Expanded} &
\textbf{Avg. Depth} & \textbf{Avg. \#Changes} \\
\midrule
\multirow{4}{*}{Entropy}
 & Depth-first  & 125.00 & \textbf{0.05} & 1.00 & 19.14  & \textbf{3.36} & 2.27 \\
 & Depth-best   & 90.91  & 0.08 & 1.00 & 37.00  & 4.32 & 1.41 \\
 & Change-first & \textbf{88.64}  &\textbf{ 0.05 }& 1.00 & \textbf{13.45 } & 4.36 & \textbf{1.36} \\
 & Change-best  & \textbf{88.64}  & 0.14 & 1.00 & 42.95  & 4.36 & \textbf{1.36 }\\
\midrule
\multirow{4}{*}{High-Cost-First}
 & Depth-first  & 123.64 & \textbf{0.04} & 1.00 & 16.73  & \textbf{3.27} & 2.27 \\
 & Depth-best   & \textbf{88.64}  & 0.10 & 1.00 & 32.50  & 3.68 & \textbf{1.36} \\
 & Change-first & \textbf{88.64 } & \textbf{0.04} & 1.00 & \textbf{9.86 }  & 3.68 & \textbf{1.36} \\
 & Change-best  & \textbf{88.64}  & 0.11 & 1.00 & 35.91  & 3.68 & \textbf{1.36} \\
\bottomrule
\end{tabular}
\end{table*}

\subsection{EXTree Construction Evaluation}
\label{subsec:EXtree-eval}


Building on the findings from the previous sub-section, we next evaluate how different policy tree constructions influence feedback efficiency and explanation quality. The \texttt{change\_first} strategy is held constant, and the optimal $(\text{max\_depth}, \text{max\_change})$ values from Table~\ref{tab:grid-optima} are applied to each tree type.

\subsubsection*{Experimental Setup}
\label{subsubsec:poltree-expsetup}

Three tree construction methods are analyzed:
\begin{itemize}
    \item \textbf{Entropy-based:} At each node, attributes with the highest entropy are selected for splitting, following the standard information-theoretic principles.
    \item \textbf{High-cost-first:} Attributes with high modification costs are prioritized closer to the root node.
    \item \textbf{Low-cost-first:} Attributes with low modification costs are prioritized closer to the root node.
\end{itemize}

The same datasets and denied requests from the previous section are used to ensure consistency and enable direct comparison.

\subsubsection*{EXTree Construction Results}
\label{subsubsec:poltree-results}

Table~\ref{tab:tree_comparison} presents performance metrics for each tree construction method across the three datasets. Table~\ref{tab:tree-results-summary} compares three tree construction strategies—Entropy-based, High-Cost-First, and Low-Change-Cost—across all datasets using the change-first feedback strategy. The results show that different heuristics lead to different trade-offs in efficiency and solution quality.

\begin{table*}[h!]
\centering
\footnotesize
\caption{Comparative performance of tree construction heuristics.}
\label{tab:tree_comparison}
\renewcommand{\arraystretch}{1.2}
\begin{tabularx}{\textwidth}{
l l
>{\raggedleft\arraybackslash}X
>{\raggedleft\arraybackslash}X
>{\raggedleft\arraybackslash}X
>{\raggedleft\arraybackslash}X
>{\raggedleft\arraybackslash}X
>{\raggedleft\arraybackslash}X}
\toprule
\textbf{Dataset} & \textbf{Tree Type} &
\textbf{Avg. Cost} & \textbf{Time (ms)} &
\textbf{Success} & \textbf{Nodes} &
\textbf{Depth} & \textbf{Changes} \\
\midrule
\multirow{3}{*}{\shortstack[l]{Synthetic-1\\(4, 30)}}
 & Entropy             & -- & -- & 0.00 & -- & -- & -- \\
 & High-Cost-First     & 133.78 & \textbf{1.23} & \textbf{0.95} & \textbf{270.55} & \textbf{14.05} & \textbf{3.64} \\
 & Low-Cost-First      & \textbf{107.07} & 7.57 & 0.93 & 1853.82 & 15.56 & 3.66 \\
\midrule
\multirow{3}{*}{\shortstack[l]{Synthetic-2\\(8, 30)}}
 & Entropy             & 445.59 & 23.77 & 0.80 & 5582.13 & 17.41 & 7.19 \\
 & High-Cost-First     & 210.53 & \textbf{2.22}  & \textbf{1.00} & \textbf{445.23}  & \textbf{14.43} & \textbf{3.78} \\
 & Low-Cost-First      & \textbf{200.10} & 10.90 & \textbf{1.00} & 2308.55 & 15.43 & 3.98 \\
\midrule
\multirow{3}{*}{\shortstack[l]{Healthcare\\(3, 5)}}
 & Entropy             & 88.64 & 0.05 & \textbf{1.00} & 13.45 & 4.36 & \textbf{1.36} \\
 & High-Cost-First     & 88.64 & \textbf{0.04} & \textbf{1.00} & \textbf{9.86 } & 3.68 & \textbf{1.36} \\
 & Low-Cost-First      & \textbf{75.45} & 0.13 & \textbf{1.00} & 22.86 & \textbf{3.55 }& \textbf{1.00} \\
\bottomrule
\end{tabularx}
\end{table*}

\begin{table}[h!]
\footnotesize
\centering
\caption{\footnotesize Aggregate performance comparison of policy tree construction heuristics across all datasets using the change-first strategy. Normalized scores are computed as the ratio of best cost to actual cost within each dataset (higher is better). Metrics are averaged only over datasets where solutions were found, and success is reported separately.}
\label{tab:tree-results-summary}
\renewcommand{\arraystretch}{1.4}
\begin{tabular}{l c c c c}
\toprule
\textbf{Tree Type} & \textbf{Norm. Score} & \textbf{Time (ms)} & \textbf{Depth} & \textbf{Success} \\
\midrule
Entropy-based & 0.65 & 11.91 & 10.89 & 0.60 \\
High-Cost-First & 0.87 & \textbf{1.23} & \textbf{10.72} & \textbf{0.98} \\
Low-Cost-First & \textbf{0.94} & 6.20 & 11.51 & \textbf{0.98} \\
\bottomrule
\end{tabular}
\end{table}

\textbf{High-Cost-First} trees achieve the best overall balance of efficiency and effectiveness. By prioritizing attributes with higher modification costs, they eliminate expensive solution branches early, creating a natural pruning effect. This leads to fewest node expansions (9.86–445.23) and the lowest runtimes (0.04–2.22~ms), while keeping the search depth moderate (3.68–14.43). Although these trees do not always produce the absolute lowest-cost policies, they consistently reach near-optimal solutions quickly.

\textbf{Low-Change-Cost} trees adopt the opposite approach. Because inexpensive attributes are explored first, the search tends to move deeper and broader into the tree before finding a feasible correction. In some cases this slightly improves solution cost—for example, on Synthetic-2 the average cost drops to 200.10 compared to 210.53 for High-Cost-First. However, the improvement is small and comes with much higher computation, requiring roughly five times more node expansions and longer runtimes.

\textbf{Entropy-based} trees perform the weakest overall. Since splits are chosen based on information gain rather than modification cost, the resulting tree structure is better suited for classification than for cost-aware search. As a result, the search expands more nodes, produces higher-cost policies, and tends to generate deeper solution paths (e.g., 17.41 compared to 14.43 and 15.43 on Synthetic-2). The search therefore spreads across many branches without consistently moving toward low-cost corrections.

\subsubsection*{Intuitive Explanation}
\label{subsubsec:illustrative_explan}

\begin{figure*}[htbp]
    \centering
    \begin{subfigure}{0.45\textwidth}
    \includegraphics[width=\linewidth]{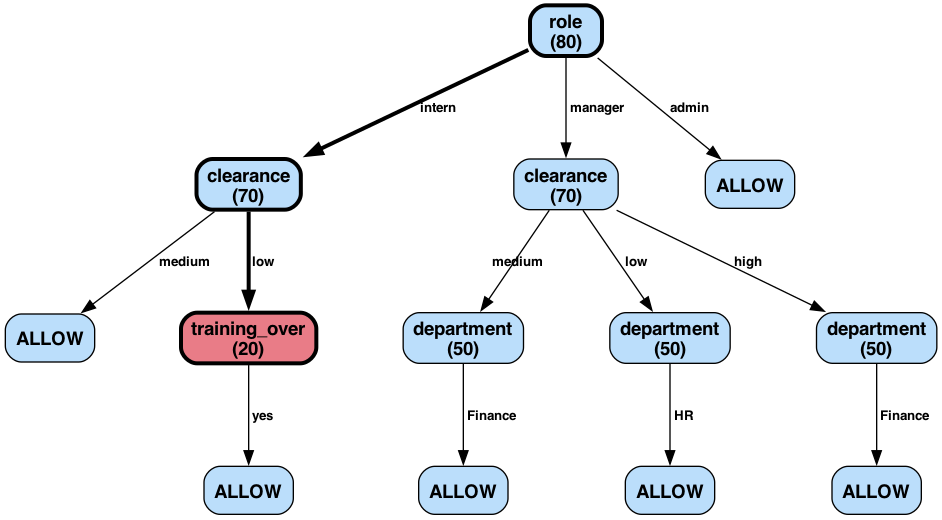}
    \caption{\footnotesize High-change-cost-first EXTree (evaluation).}
    \label{fig:highfirst}
    \end{subfigure}
    \begin{subfigure}{0.45\textwidth}
   \includegraphics[width=\linewidth]{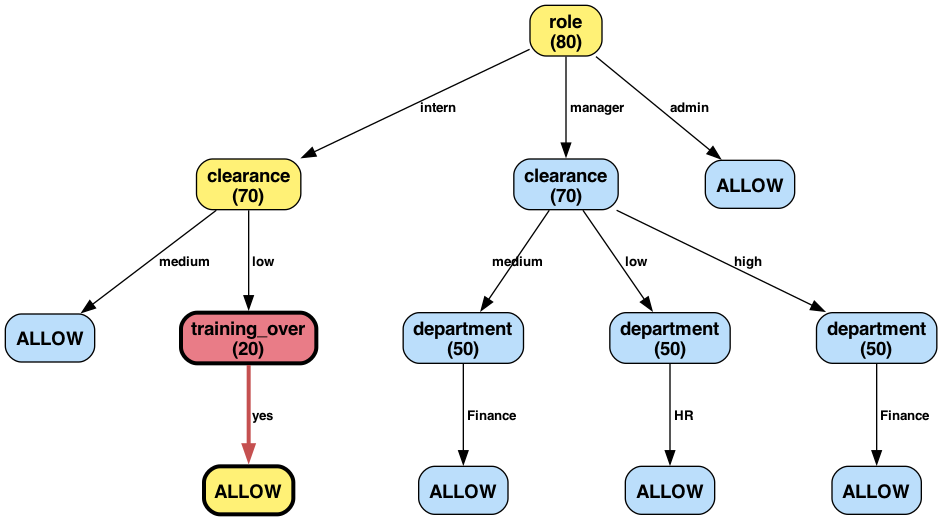}
       \caption{\footnotesize High-change-cost-first EXTree (feedback).}
    \label{fig:changefirst_high}
    \end{subfigure}
    \begin{subfigure}{0.45\textwidth}
    \includegraphics[width=\linewidth]{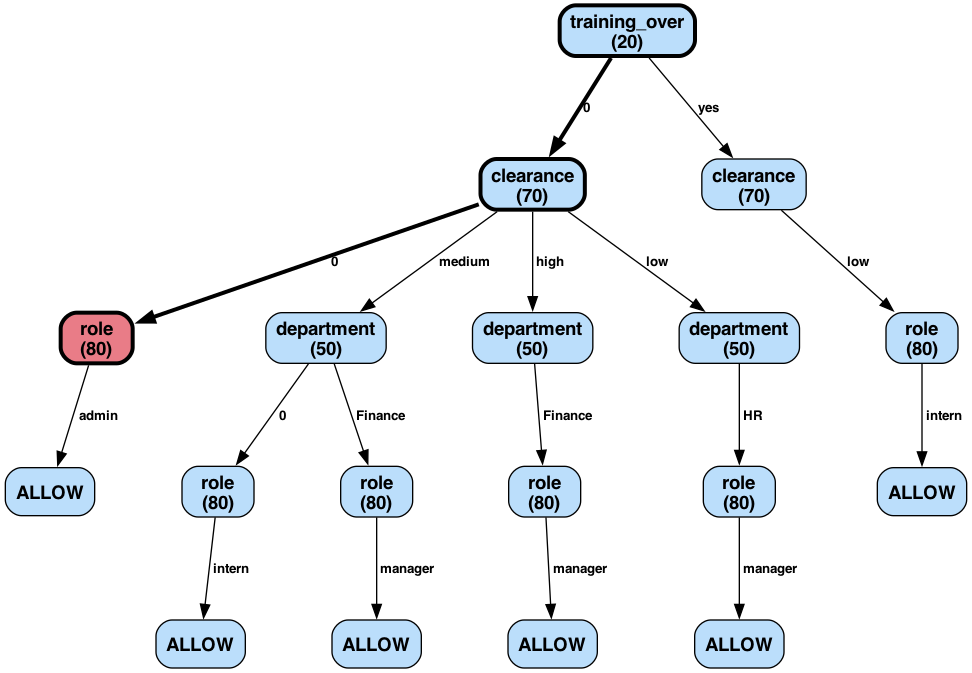}
    \caption{\footnotesize Low-change-cost-first EXTree (evaluation).}
    \label{fig:lowfirst}
\end{subfigure}
\begin{subfigure}{0.45\textwidth}
\includegraphics[width=\linewidth]{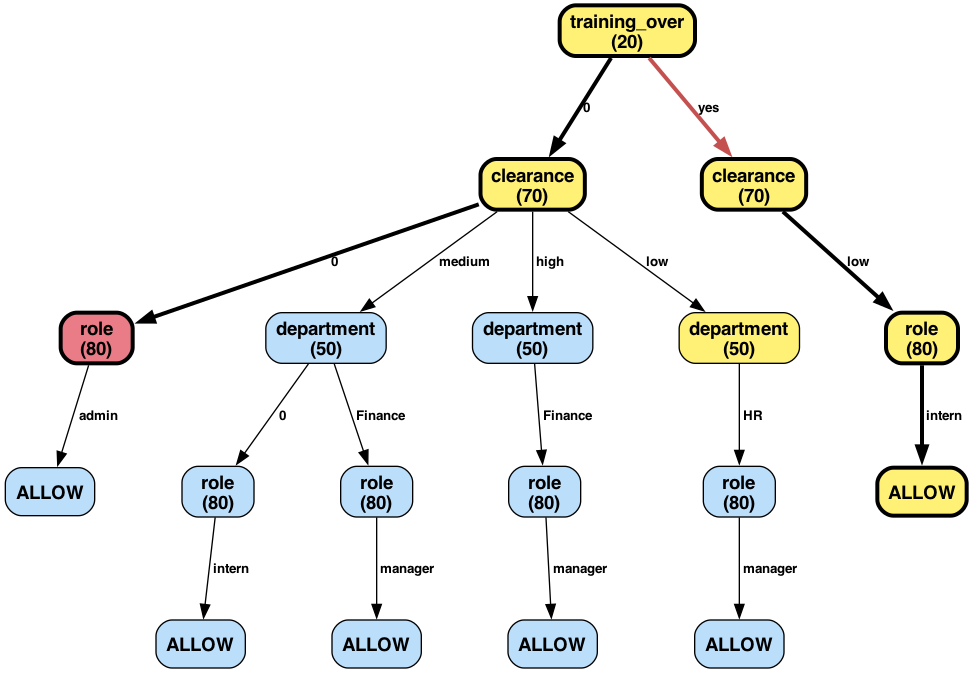}
    \caption{\footnotesize Low-change-cost-first EXTree (feedback).}
    \label{fig:changefirst_low}
        \end{subfigure}
    \caption{\footnotesize Comparison of EXTree construction strategies using the policy introduced in Section~\ref{subsec:prelims_abac} and the denied request 
$(\texttt{role}=\texttt{intern}, \texttt{department}=\texttt{General}, \texttt{clearance}=\texttt{low}, \texttt{training\_over}=\texttt{no})$. 
Although both trees have identical depth, their construction significantly affects the amount of search required to generate actionable feedback.}

\label{fig:exampleEXTreeXonstruction}
\end{figure*}

Figures~\ref{fig:highfirst} and~\ref{fig:lowfirst} show two EXTrees constructed from the same policy introduced in Section~\ref{subsec:prelims_abac}. In the \emph{high-change-cost-first} construction, attributes that are expensive or difficult to modify appear closer to the root, whereas attributes with lower changeability cost are placed deeper in the tree. In contrast, the \emph{low-change-cost-first} construction places easily modifiable attributes near the root.
The request considered here is
$
r = (\text{role} = \text{intern},\ \text{department} = \text{General},\ \text{clearance} = \text{low},\ \text{training\_over} = \text{no}),
$
which evaluates to a \textit{deny} decision under the policy.
Although the cost of the path from the deny node to an allow leaf is identical for both trees, the number of nodes traversed differs significantly.

The high-change-cost-first EXTree (Figure~\ref{fig:highfirst}) requires traversing fewer nodes to locate an appropriate allow node, even though both trees have the same depth. This difference arises because attributes that are more \textit{changeable} (i.e., lower change cost) are intentionally positioned closer to the leaves, where denials are more likely to occur.  
Consequently, once a denial is encountered, the subsequent local search is confined to attributes that the requester can modify feasibly, leading to rapid generation of meaningful feedback (Figure \ref{fig:changefirst_high}).

In contrast, when attributes with low change cost are placed near the root (Figure~\ref{fig:lowfirst}), the evaluation path for most requests is determined by easily modifiable attributes early on. This results in deeper subtrees dominated by high-cost or immutable attributes, which not only reduce the relevance of the branch to the request but also make corrective feedback harder to generate  (Figure \ref{fig:changefirst_low}).

In general, placing high-change-cost attributes near the EXTree root yields two distinct advantages. (i) 
\textbf{Branch relevance:} Early splits occur on stable, user-invariant attributes (e.g., role, department), ensuring that the evaluation path corresponds closely to the requester’s fundamental context and (ii) \textbf{Feedback efficiency:} The remaining attributes along the path from a \textit{deny} node to the nearest \textit{allow} node are those that are most easily changeable, reducing both traversal effort and overall change cost.

Therefore, EXTrees constructed using a high-change-cost-first criterion naturally align with the goal of producing low-cost, relevant feedback. This design balances efficiency with explainability.

\begin{table}[t]
\caption{Average impact of decreasing visibility on feedback generation performance in Synthetic-1 Dataset, aggregated across all search strategies.}
\label{tab:visibility-avg}
\centering
\small
\begin{tabular}{c c c c c}
\toprule
Visibility & Found (\%) & Avg Score & Avg Nodes & Avg Time (ms)\\
\midrule
100\% & 100 & 186.7 & 2478 & 10.2\\
90\%  & 100 & 205.1 & 2268 & 7.0\\
80\%  & 100 & 203.1 & 750  & 1.8\\
70\%  & 82  & 255.4 & 399  & 1.0\\
60\%  & 59  & 284.9 & 137  & 0.4\\
50\%  & 53  & 301.0 & 556  & 1.4 \\
40\%  & 13  & 301.3 & 226  & 0.6\\
$\leq$30\% & 0 & -- & 0 & 0\\
\bottomrule
\end{tabular}
\end{table}
\subsection{Effect of Visibility Constraints}
We evaluate the impact of visibility-aware filtering by restricting access to attribute--value predicates, while keeping the underlying EXTree unchanged. Visibility is modeled as a binary property of attribute--value pairs. For this experiment, the same visibility assignment is applied uniformly across all users to isolate the effect of restricted disclosure. We increase the fraction of hidden attribute--value predicates across the policy. During feedback generation, paths traversing hidden predicates are rendered infeasible by assigning infinite cost.

Table~\ref{tab:visibility-avg} summarizes the effect of decreasing visibility on feedback generation for the Synthetic-1 dataset. As visibility decreases, the fraction of requests for which a valid explanation exists drops sharply; below approximately 30\% visibility, no feasible explanation paths remain. Similar qualitative trends are observed across other datasets.

Crucially, visibility constraints introduce no measurable computational overhead. Average feedback generation time remains stable---and often decreases slightly---as visibility is reduced from 100\% to 50\%, remaining within sub-millisecond bounds. This is explained by the early exclusion of infeasible paths via infinite traversal cost.

Average explanation cost increases as visibility decreases, reflecting the elimination of lower-cost paths that traverse hidden predicates and the selection of higher-cost feasible alternatives. Despite reduced feasibility, node expansion and traversal depth remain bounded until explanations disappear entirely. Find an illustrative example in Appendix~\ref{vis_ex}.

\begin{table}[t]
\centering
\caption{Comparison of ABAC systems with respect to Explainability Support.}
\label{tab:related_comparison}
\resizebox{1.0\columnwidth}{!}{
\begin{tabular}{lccccc}
\toprule
System 
& Denial 
Explanation 
& Actionable 
Feedback 
& Scalability 
& Privacy 
Control 
& Explanation 
Mechanism \\
\midrule

KNOW~\cite{kapadia2004know} 
& \checkmark 
& \checkmark 
& $\times$ 
& \checkmark 
& OBDD-based reasoning \\

CUE~\cite{cue_framework} 
& \checkmark 
& \checkmark 
& $\times$ 
& \checkmark 
& XACML policy analysis \\

PolTree~\cite{nath2019poltree} 
& $\times$ 
& $\times$ 
& \checkmark  
& $\times$ 
& Hierarchical policy tree \\

Margrave~\cite{margrave} 
& $\times$ 
& $\times$ 
& $\times$ 
& $\times$ 
& Decision-diagram verification \\

LLMAC~\cite{llmac_2026} 
& \checkmark 
& $\times$ 
& $\times$ 
& $\times$ 
& LLM-based explanation \\

\textbf{EXTree (This Work)} 
& \checkmark 
& \checkmark 
& \checkmark 
& \checkmark 
& Policy tree + bounded local search \\

\bottomrule
\end{tabular}
}
\end{table}

\subsubsection*{Discussion}
\label{subsubsec:poltree_concl}

Entropy-based trees provide rich feedback options due to their balanced splits, but require deeper traversal to reach valid allow nodes. High-change-cost-first trees lead to more interpretable feedback—users receive guidance on modifying high-impact attributes early in the path. Conversely, low-change-cost-first trees offer the simplest, low-effort fixes but at the expense of feedback completeness. The combination of the \texttt{change\_first} strategy with the high-change-cost-first EXTree structure provides the most effective feedback mechanism for explainable ABAC systems producing concise and meaningful feedback while maintaining low response latency, thereby addressing RQ3.

\section{Related Work}
\label{sec:related}

With increasingly complex models being used in computer security and a growing societal demand for transparency in automated decision making, explainability has emerged as an important consideration in access control. Yet, explicit denial-feedback mechanisms remain rare in  ABAC systems.
As Miller~\cite{miller2019explanation} argues, effective explanations must be selective, contrastive, and contextually relevant. In access control, this translates to providing clear, actionable feedback rather than exposing the full policy logic.
The earliest systems to provide structured denial feedback are KNOW~\cite{kapadia2004know} and CUE~\cite{cue_framework}. KNOW computes minimal sufficient policy changes using OBDD-based reasoning and regulates disclosure through meta-policies. CUE enhances XACML infrastructures with user-oriented feedback by distinguishing mutable and immutable attributes. While both frameworks improve usability, feedback is derived through global policy traversal (decision-diagram reasoning or policy scanning), which limits scalability in large-scale deployments.

Other lines of work focus on efficient policy evaluation or formal verification rather than user-facing feedback. PolTree~\cite{nath2019poltree} introduces hierarchical indexing for efficient ABAC decision-making but does not support explanation. Margrave~\cite{margrave} translates policies into decision diagrams for verification and semantic differencing, producing counterexamples for administrators rather than user-facing denial explanations.

With new access control models being developed, there is renewed interest in exploring explainability in security systems more broadly~\cite{viganoexplainabilitysecurity}. Mehri et al.~\cite{explainablesacmat2025} motivate the need for explainable access control and outline key research challenges, but do not introduce a concrete feedback mechanism. Rodríguez et al.~\cite{debac} propose DEBAC, which employs explainable boosting machines to compute interpretable trust scores in WLAN environments; however, it does not compute minimal corrective attribute changes for denied access. More recently, LLM-based approaches such as LLMAC propose generating natural-language explanations for access decisions using large language models. While such systems improve interpretability at the presentation layer, explanations are derived through statistical inference rather than structured policy reasoning and do not provide bounded, minimal-change guarantees within the policy model itself.

Related work in identity management and data governance~\cite{xaidatagovernance}, explainable intrusion detection systems~\cite{neupane2022explainableintrusiondetectionsystems,exaiids1} and surveys on explainable cybersecurity~\cite{exaicybersecurity}, underscores the importance of transparency, though these efforts operate outside the ABAC denial-feedback setting.
Unlike OBDD-based or policy-scanning explanation layers, EXTree incorporates explainability directly into the hierarchical policy structure itself. Feedback generation is expressed as a bounded local search around the deny node, optimizing changeability costs while respecting visibility constraints. To the best of our knowledge, no existing ABAC framework formulates explainability as a bounded search process designed explicitly for scalable feedback generation. Table~\ref{tab:related_comparison} summarizes the key distinctions.

\section{Conclusion and Future Directions}
\label{sec:concl}
This paper demonstrates that efficiency and explainability need not be competing objectives by proposing \textit{EXTrees} for ABAC. While traditional feedback mechanisms require expensive scans of the policy after a denial, EXTrees encode attribute dependencies directly in the tree structure, allowing the system to produce interpretable feedback focused on actionable attribute changes while keeping computation lightweight. As a result, efficient policy evaluation and transparent feedback can be achieved within the same architecture.


There are several directions for future work. A natural next step is to evaluate EXTree on real-world ABAC systems, where attributes may be correlated and constraints depend on context. This would help understand practical applicability. Another direction is extending EXTree to support richer policy semantics, such as rule-combining operators (e.g., permit-overrides and deny-overrides) used in XACML. These operators could be supported by encoding precedence constraints in the tree during construction or traversal. Adaptive feedback strategies can be explored that adjust based on user behavior and attribute information, potentially using machine learning to tune system parameters over time. Finally, supporting incremental updates would allow the tree to be updated as policies change, without rebuilding it from scratch, making it more effective in dynamic environments.

We hope this work encourages further exploration of structural approaches to transparency in authorization systems.

\printbibliography

\appendix
\section{Appendix}

\subsection{Dataset Details}
\label{dataset_details}

This study employs two synthetic datasets for experimentation and one realistic dataset from the ABACLab repository~\cite{abaclab_github} to ensure real-world relevance.
The synthetic datasets were generated using the publicly available github repository of~\cite{paul2021efficient}. For simulating realistic policy overlaps, we introduced wildcards in 30–50\% of attribute values. 
\begin{table*}[h]
\centering
\caption{Summary of datasets used in evaluation. 
$n_u$: users; 
$n_o$: objects; 
$n_e$: environment entities; 
$u_a$, $o_a$, $e_a$: user/object/environment attributes; 
$nv_*$: values per attribute; 
$n_p$: policies; 
$n_{ops}$: operations or granted permissions.}
\label{tab:all_datasets}
\begin{tabular}{lccccccccccc}
\toprule
\textbf{Dataset} & $n_u$ & $n_o$ & $n_e$ & $u_a$ & $o_a$ & $e_a$ & $nv_u$ & $nv_o$ & $nv_e$ & $n_p$ & $n_{ops}$ \\
\midrule
Synthetic--1 & 1000 & 1000 & 10 & 6 & 6 & 6 & 4 & 4 & 4 & 1000 & 3 \\
Synthetic--2 & 2000 & 1500 & 20 & 8 & 8 & 6 & 5 & 5 & 4 & 2000 & 1 \\
Healthcare (ABACLab) & 21 & 16 & -- & 6 & 7 & -- & -- & -- & -- & 6 & 43 \\
\bottomrule
\end{tabular}
\end{table*}
Synthetic--1 provides a balanced baseline with 1000 users, objects, and policies, each defined by six attributes with four possible values. Synthetic--2 increases the number of entities to evaluate scalability. The parameters are chosen to produce realistic datasets while keeping the experiments manageable, avoiding cases where policies are either too sparse or too dense~\cite{yuan2005abac,hu2014guide}. Request sets maintain an approximate 70:30 accept-to-deny ratio. The Healthcare Access Dataset from ABACLab~\cite{abaclab_github,bui2025abaclab} models access control scenarios in healthcare environments, with policies defined by staff roles, patient sensitivity levels, and contextual attributes such as emergency status.
Full parameter settings are shown in Table~\ref{tab:all_datasets}.

\begin{figure}[h]
\centering

\begin{subfigure}{1\textwidth}
\centering
\includegraphics[width=0.24\linewidth]{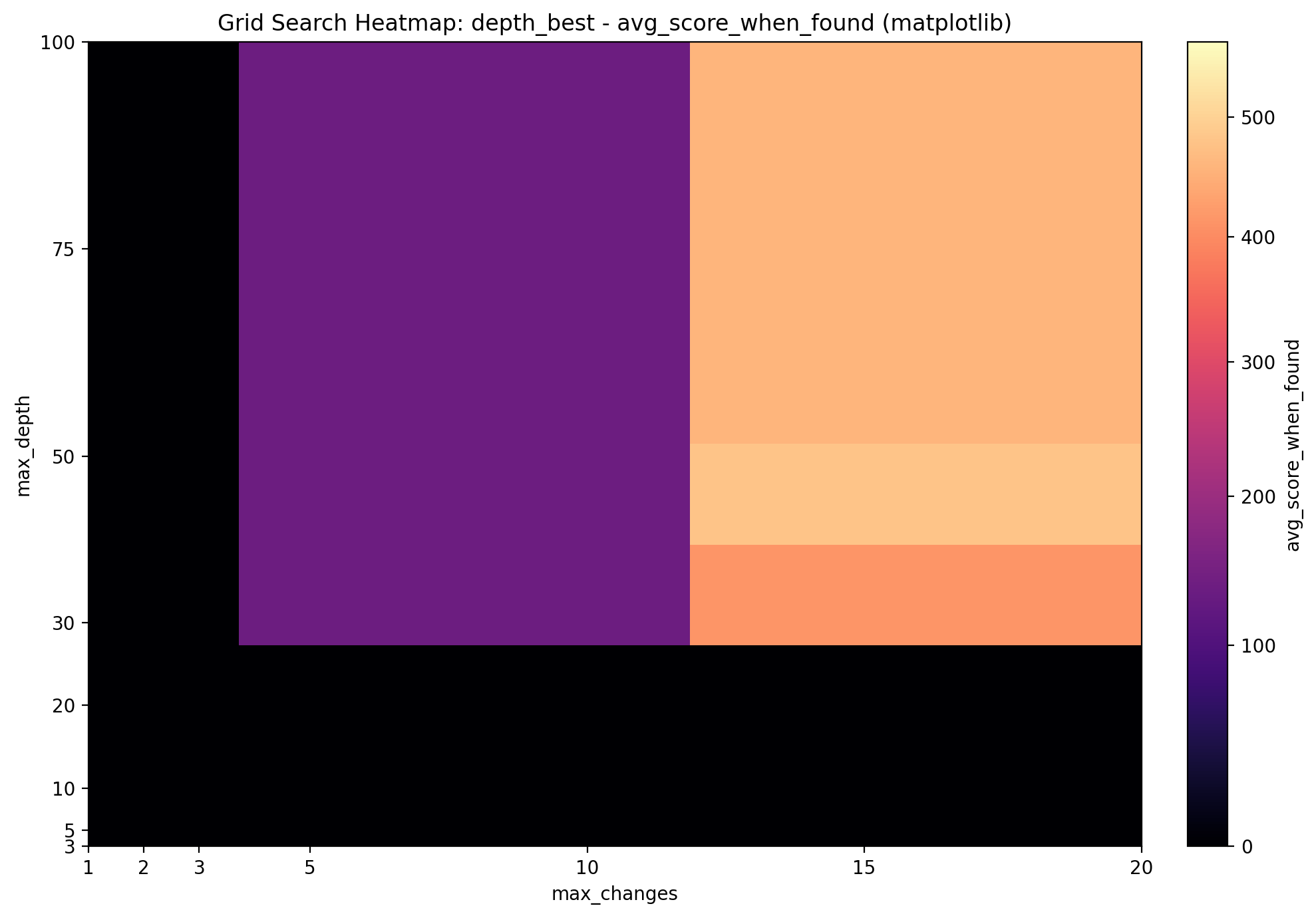}
\includegraphics[width=0.24\linewidth]{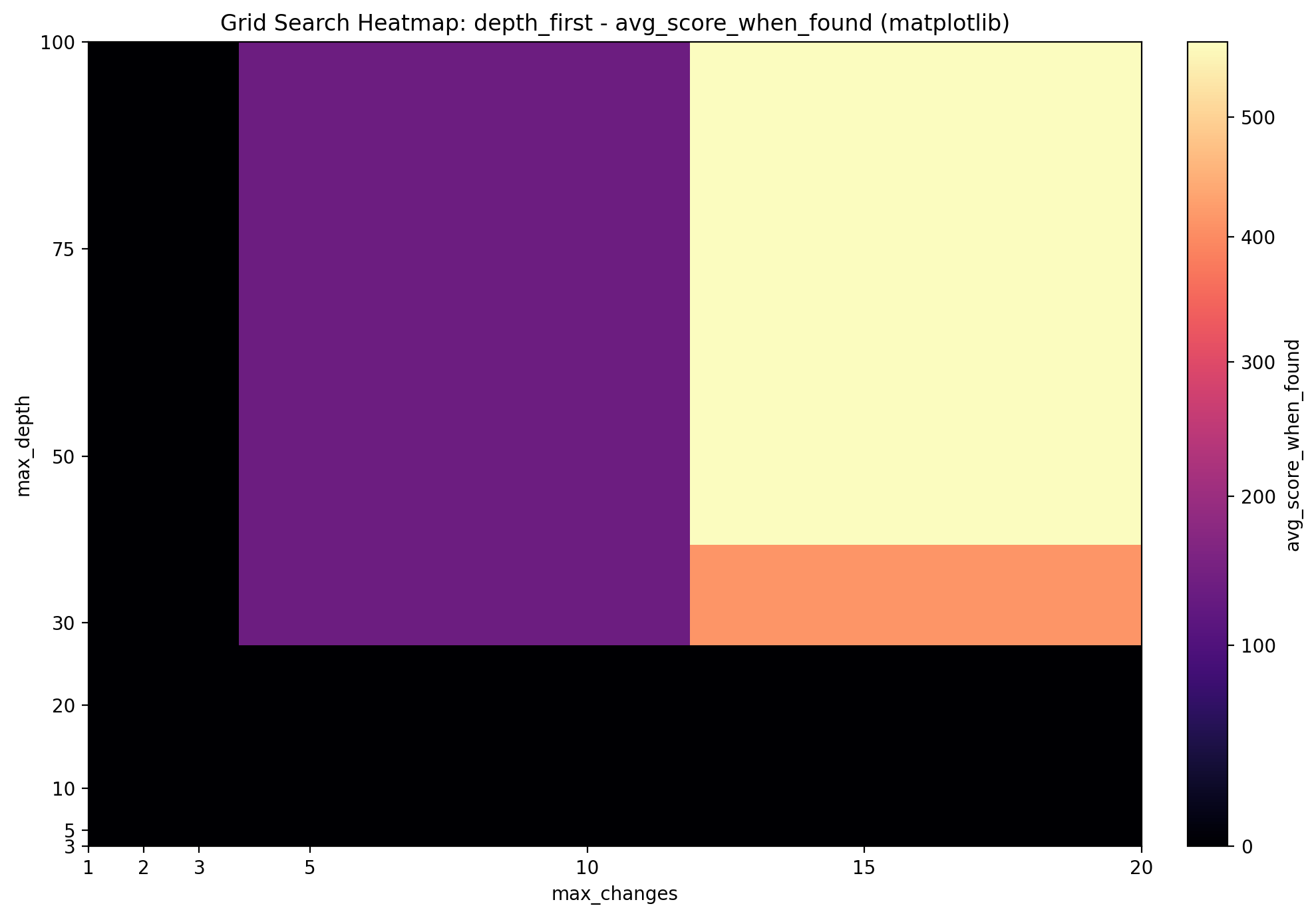}
\includegraphics[width=0.24\linewidth]{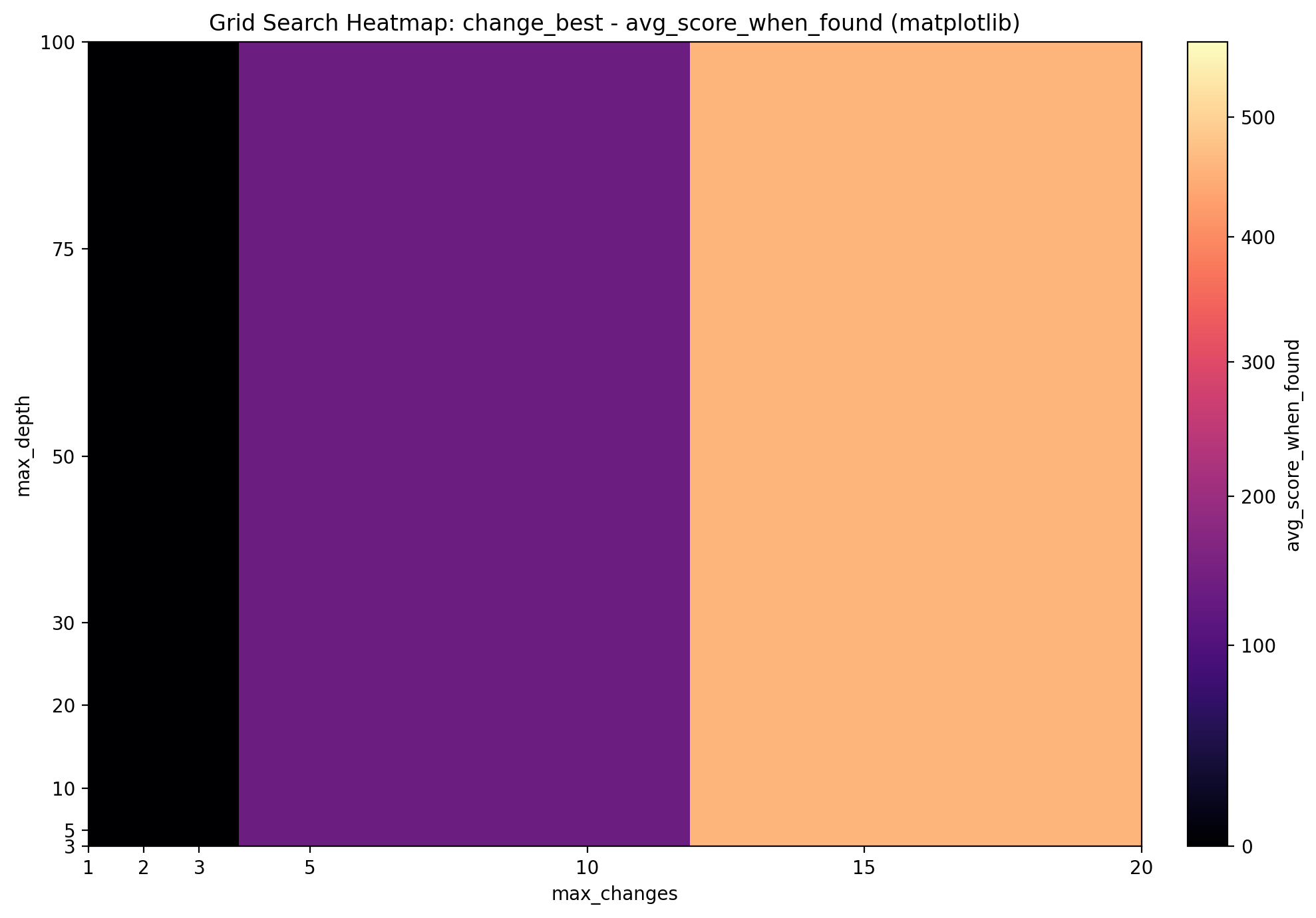}
\includegraphics[width=0.24\linewidth]{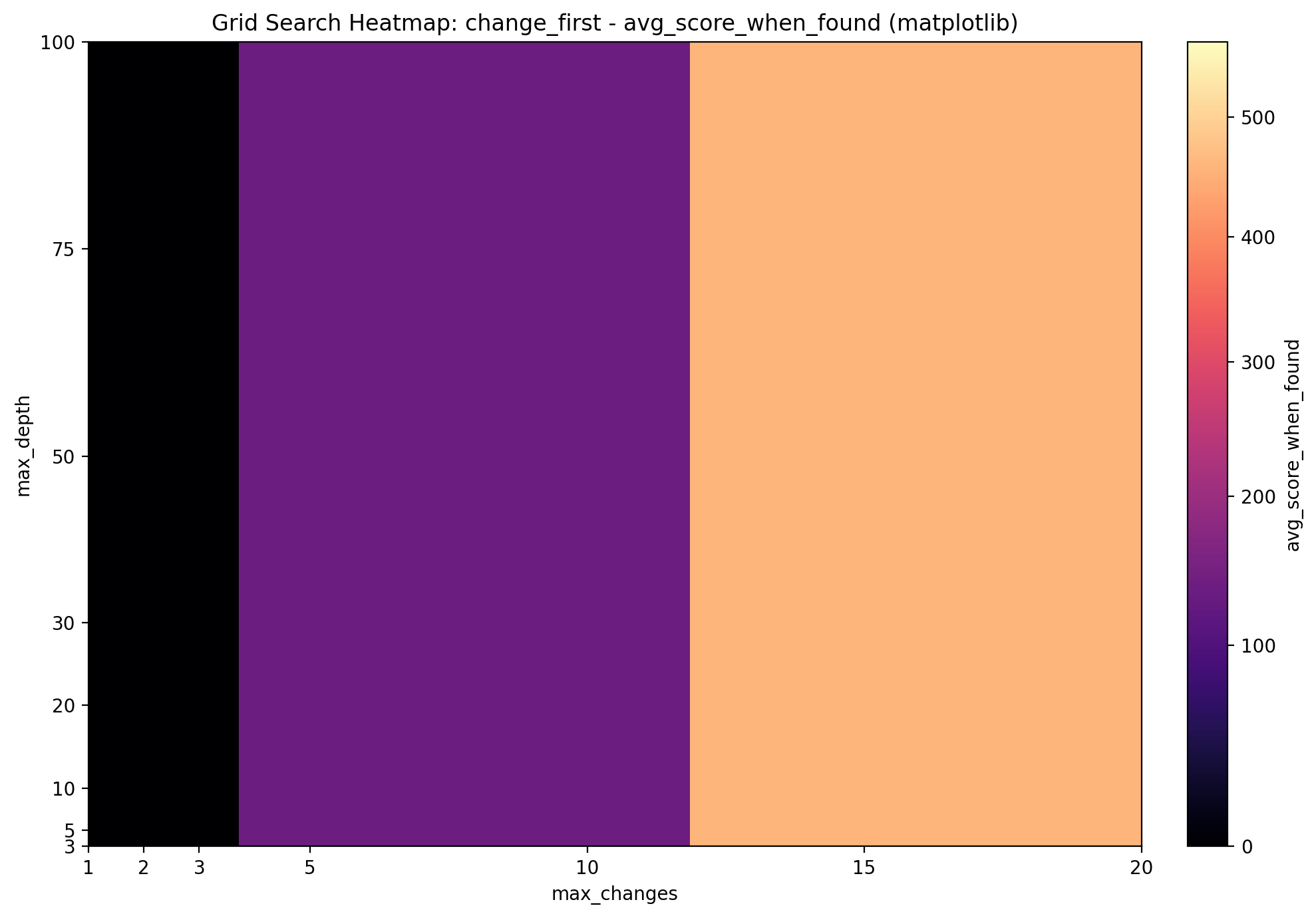}

\caption*{(a) Entropy-based}
\end{subfigure}
\hfill

\begin{subfigure}{1\textwidth}
\centering
\includegraphics[width=0.24\linewidth]{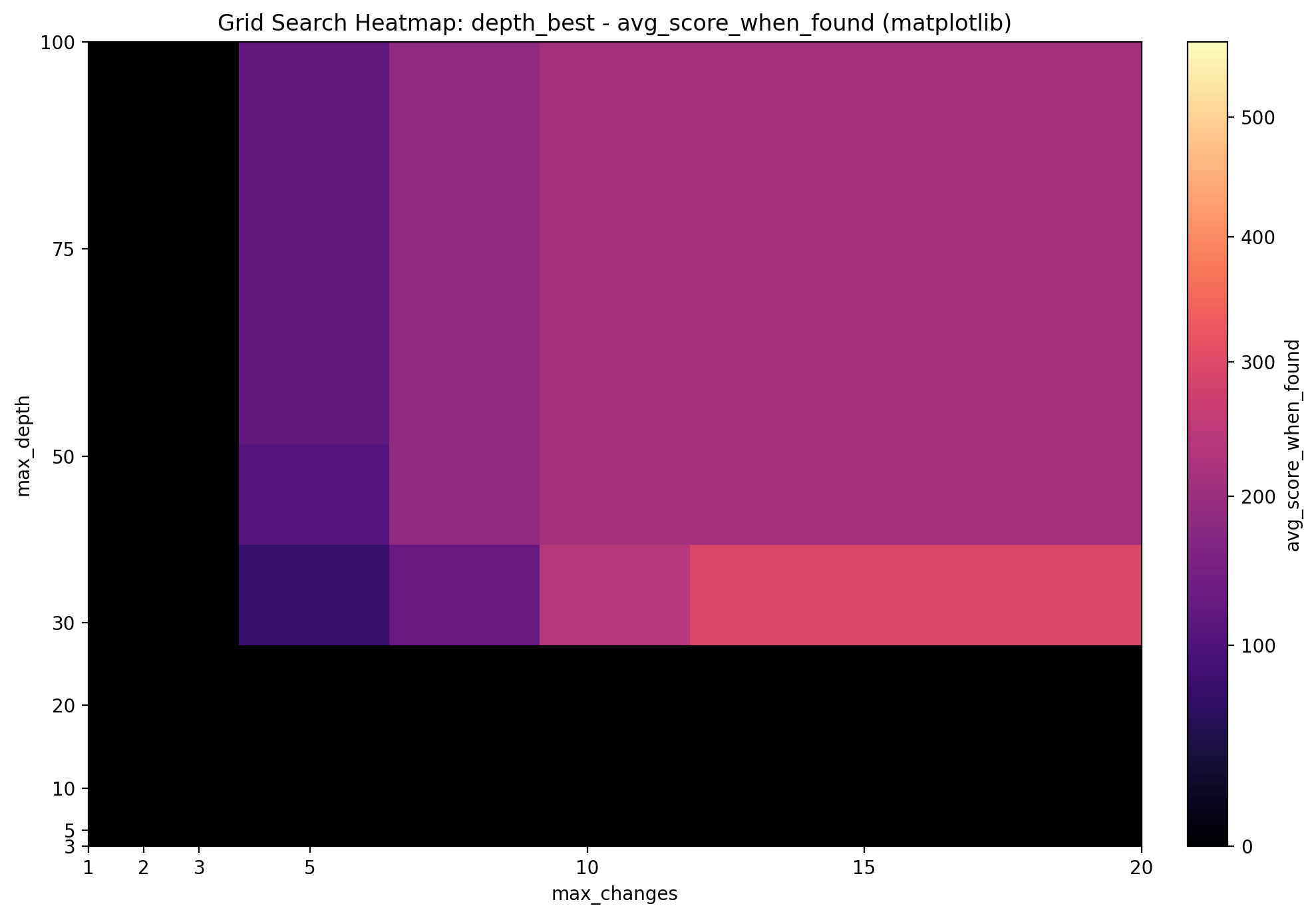}
\includegraphics[width=0.24\linewidth]{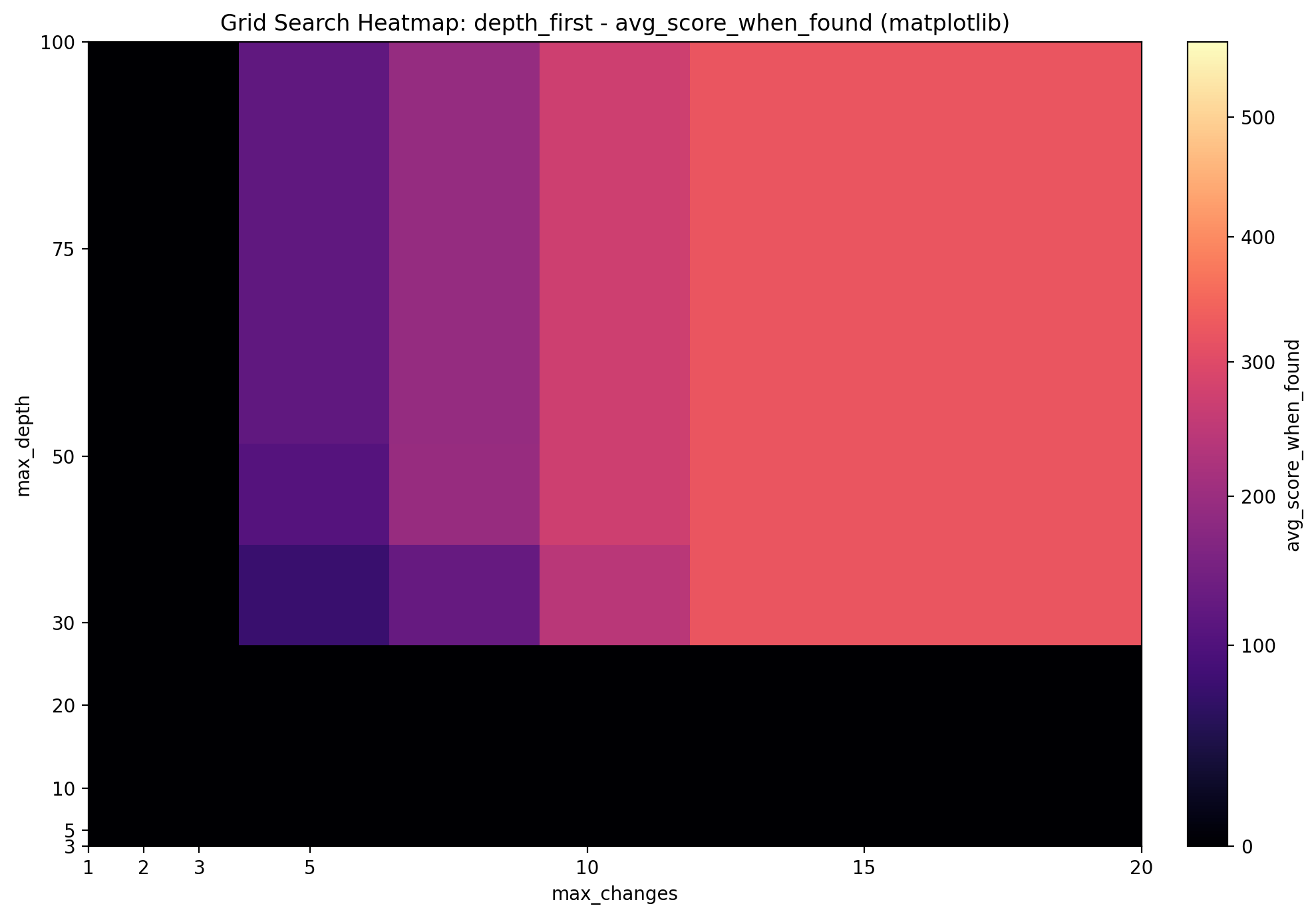}
\includegraphics[width=0.24\linewidth]{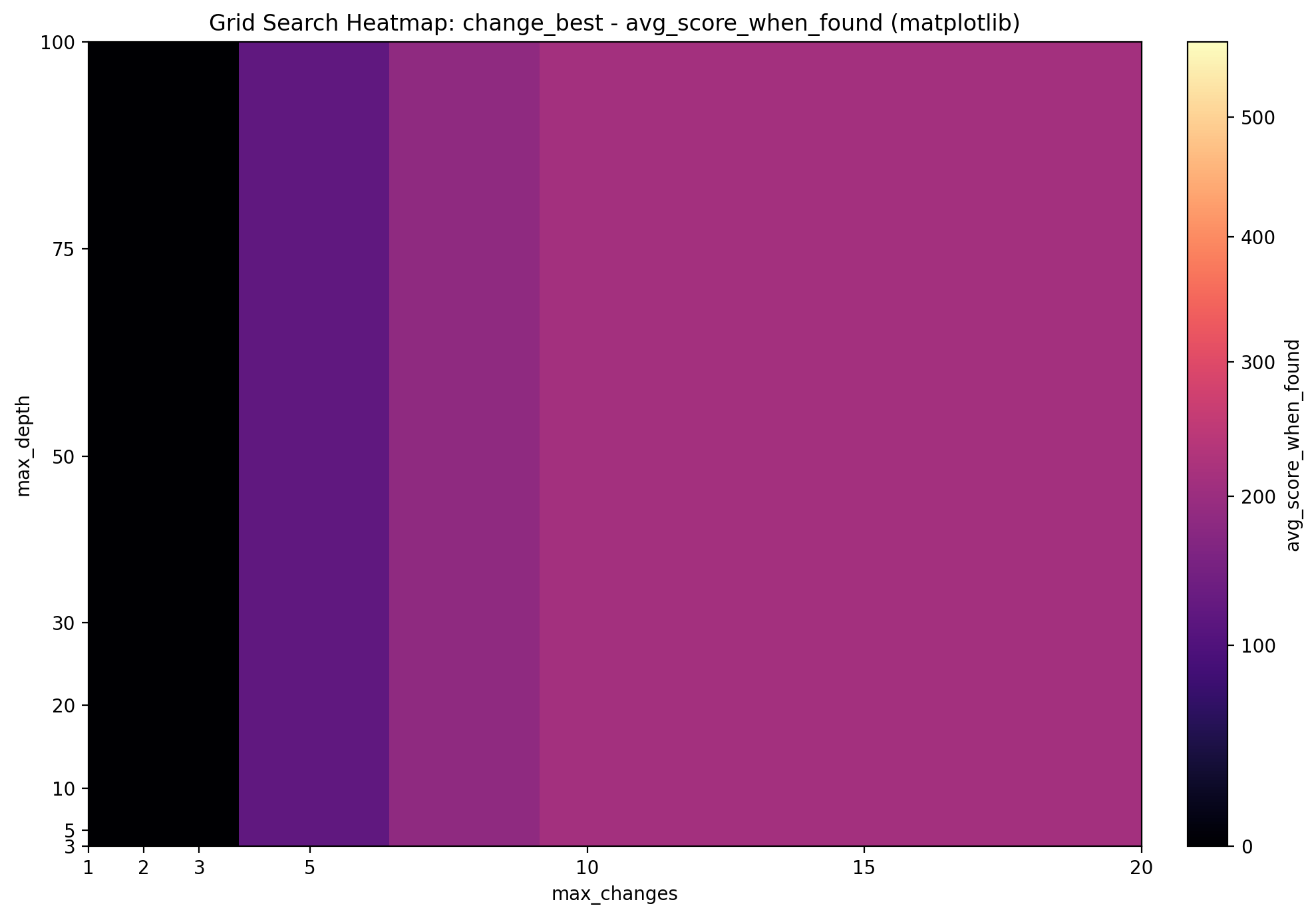}
\includegraphics[width=0.24\linewidth]{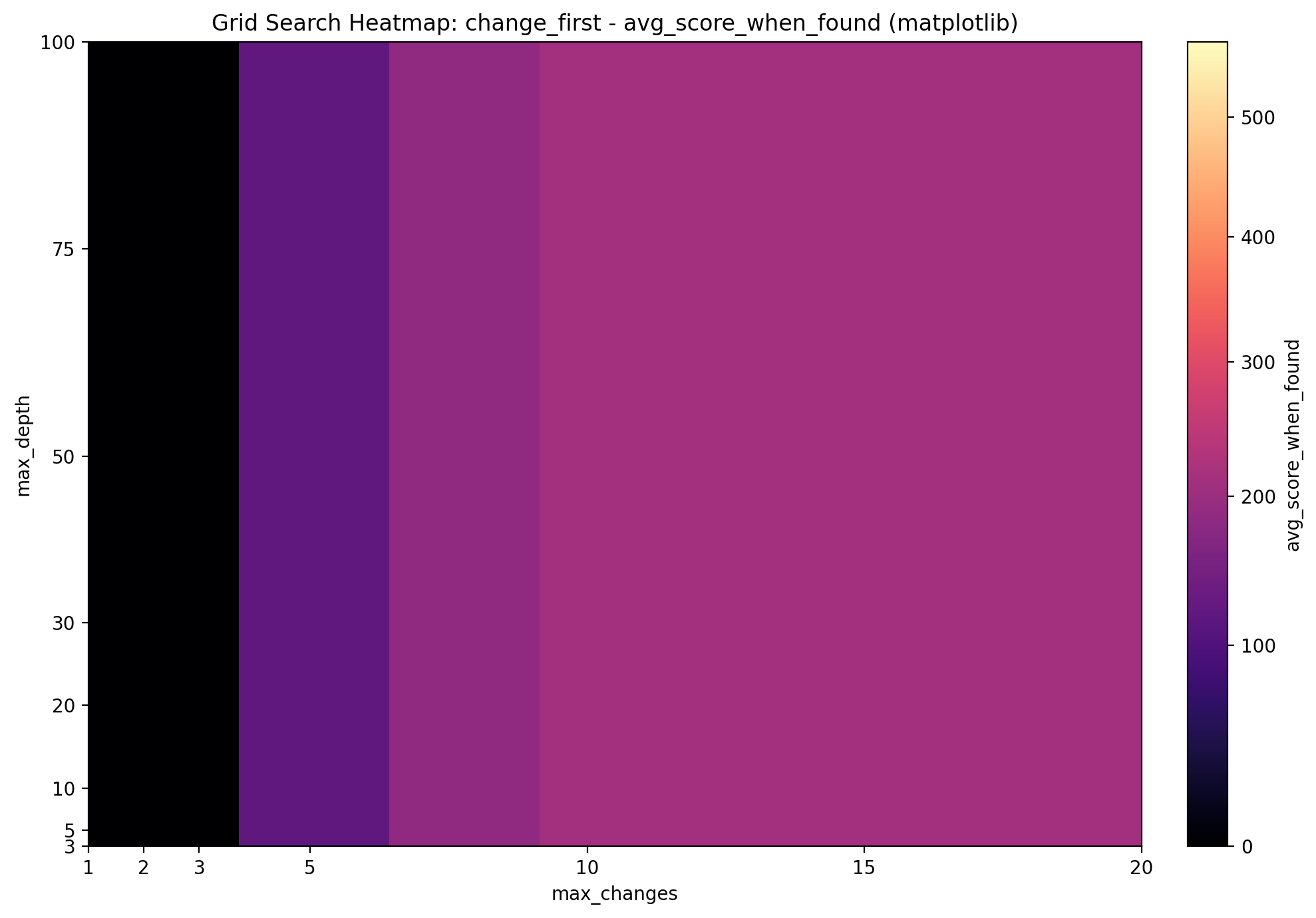}

\caption*{(b) High-change-cost-first}
\end{subfigure}
\hfill

\begin{subfigure}{1\textwidth}
\centering
\includegraphics[width=0.24\linewidth]{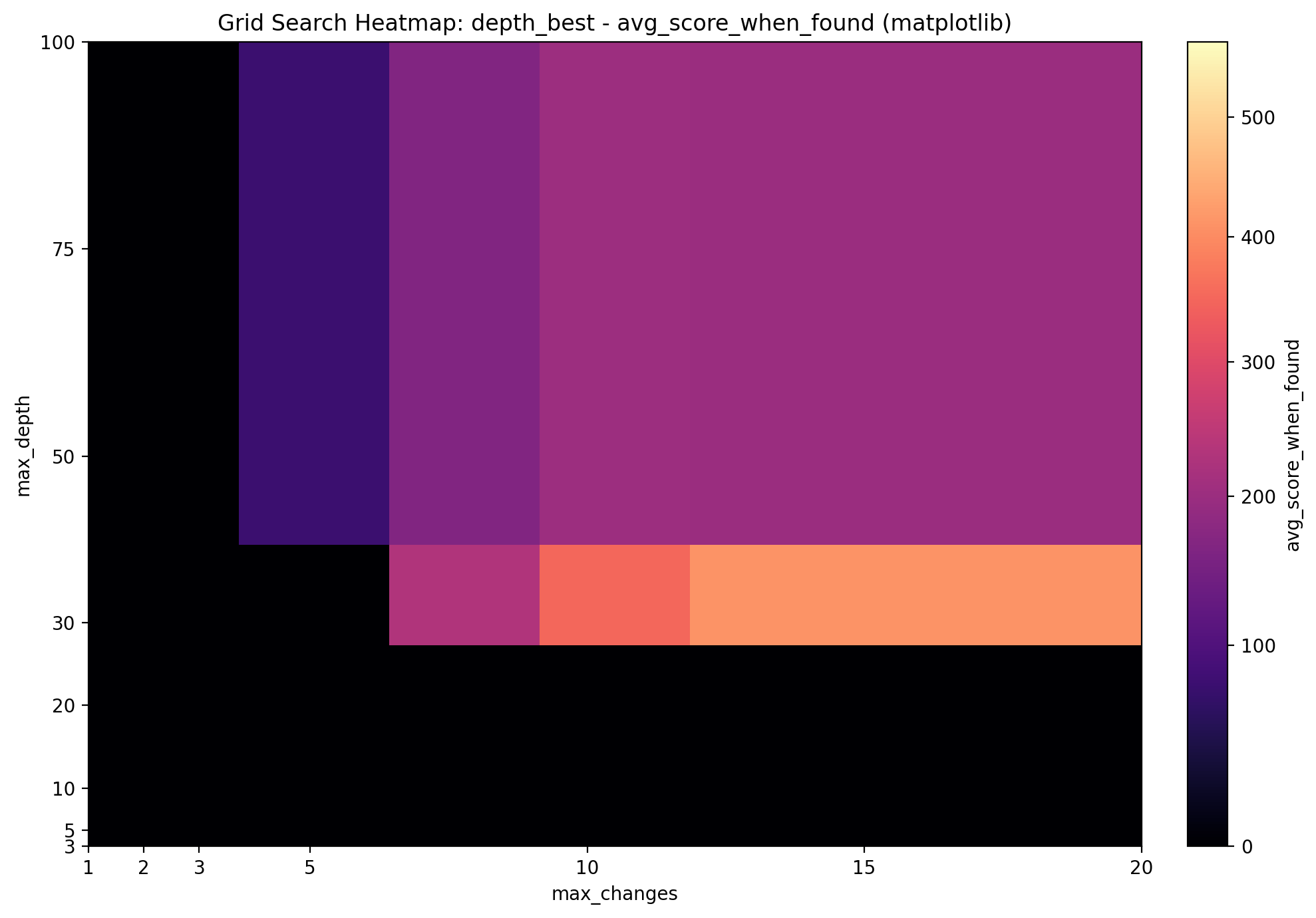}
\includegraphics[width=0.24\linewidth]{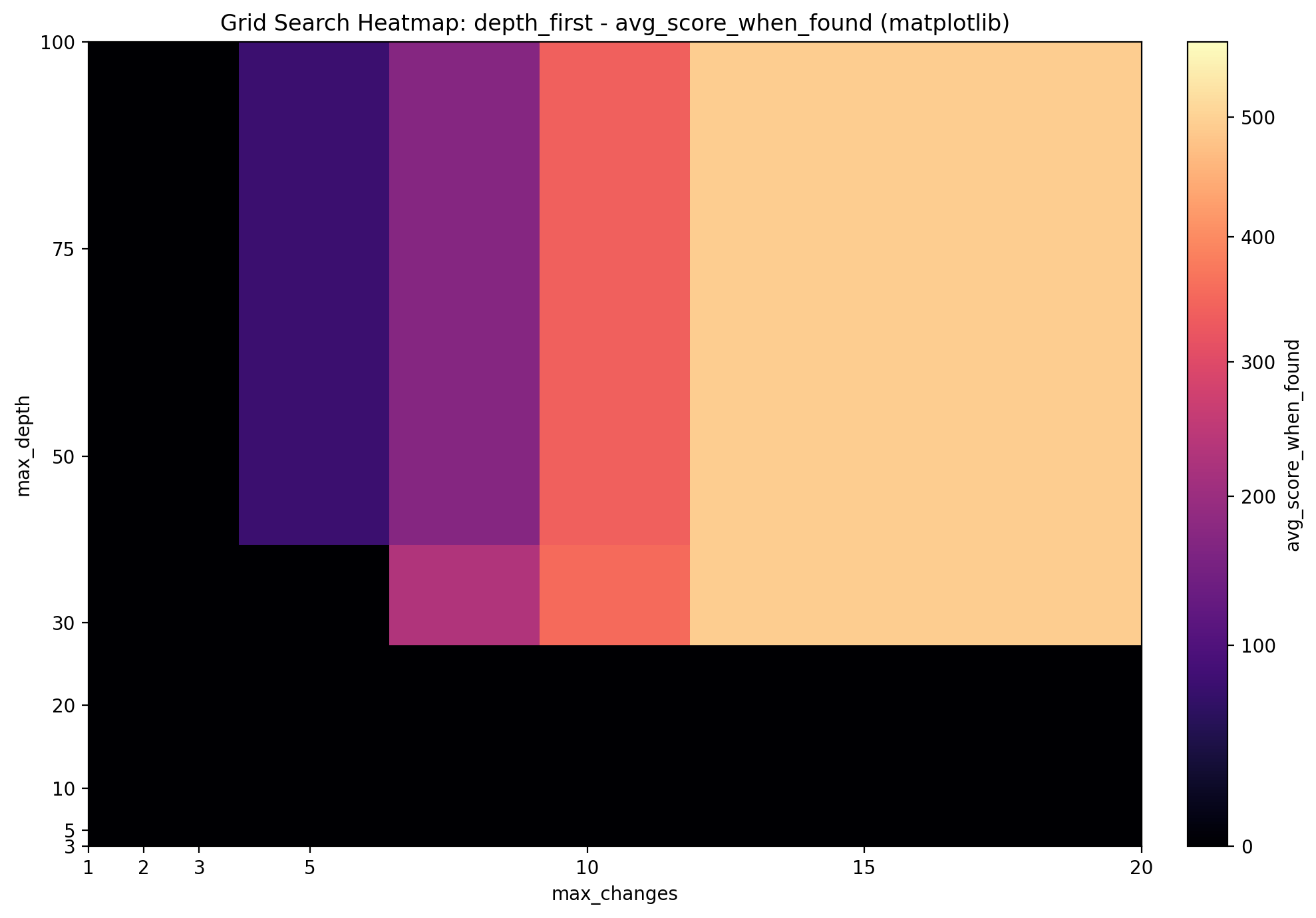}
\includegraphics[width=0.24\linewidth]{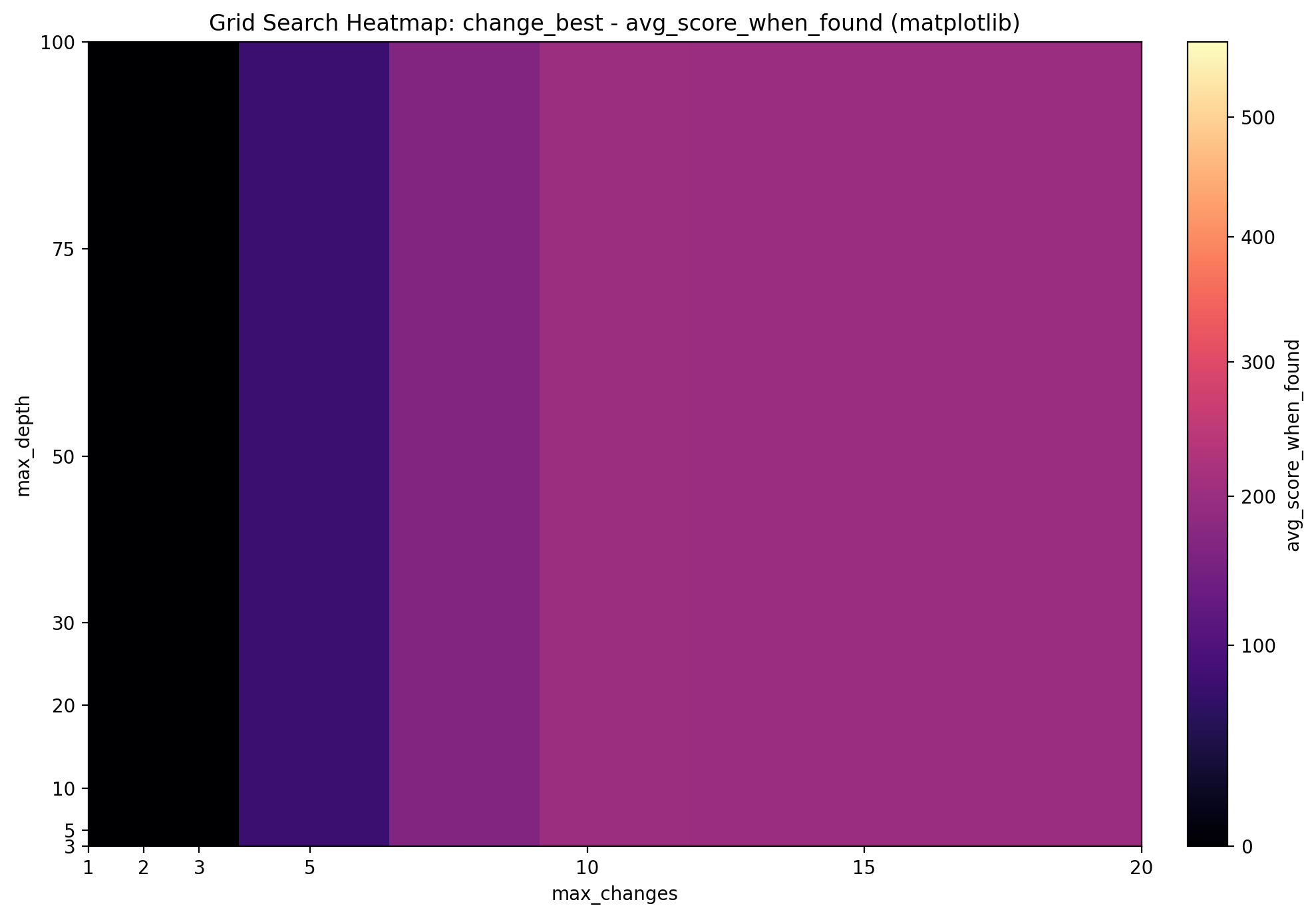}
\includegraphics[width=0.24\linewidth]{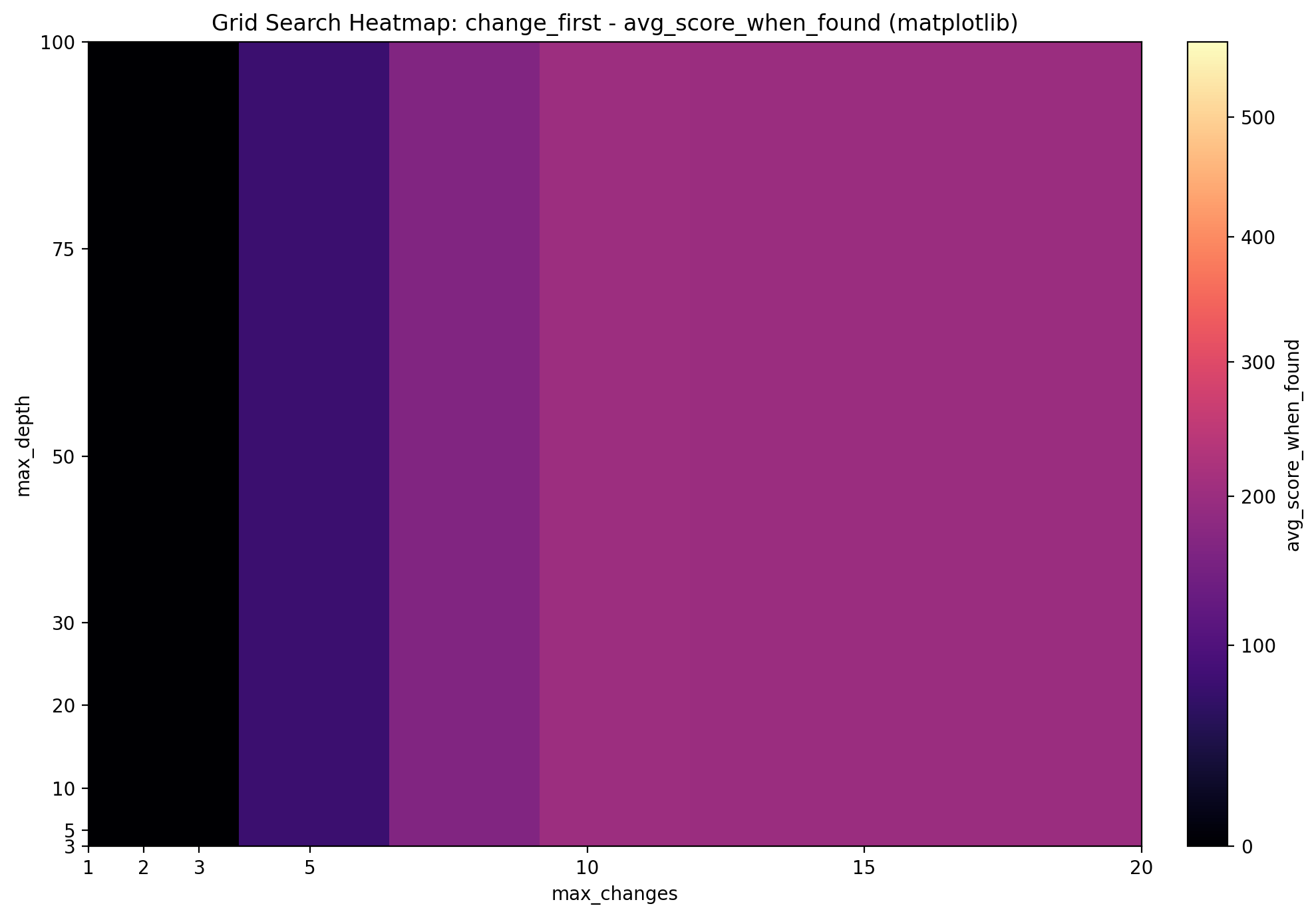}

\caption*{(c) Low-change-cost-first}
\end{subfigure}

\caption{
Cumulative feedback cost heatmaps grouped by tree construction strategy.
Within each block, the four feedback strategies (top-left to bottom-right) are: depth-best, depth-first, change-best, and change-first.
}
\label{fig:grouped_heatmaps}
\end{figure}

\subsection{Parameter Optimization}
\label{subsubsec:paramopt}

We perform a systematic grid search for each dataset over \texttt{max\_depth} and \texttt{max\_changes} to analyze their effect. 
Figure~\ref{fig:grouped_heatmaps} illustrates this process for the Synthetic-2 dataset.
The grid search varies
\texttt{max\_depth} $\in \{3, 5, 10, 20, 30, 50, 75, 100\}$ and
\texttt{max\_changes} $\in \{1, 2, 3,\\ 5, 10, 15, 20\}$.
Each heatmap cell represents a specific (\texttt{max\_depth}, \texttt{max\_changes}) configuration, where color intensity indicates the average cumulative changeability cost of the feedback when a solution is found. Black regions correspond to parameter settings where no actionable feedback is generated, while dark regions indicate simpler and more interpretable feedback.
Comparing tree constructions, entropy-based trees produce higher costs and larger infeasible regions, whereas change-aware constructions reduce both feedback cost and the number of infeasible configurations. Among feedback strategies, change-based search consistently achieves lower costs and smaller infeasible regions than depth-based variants, suggesting better coverage and feedback quality under similar parameters.
As modification limits increase, all strategies are able to resolve progressively harder denied requests that require more complex attribute changes, which naturally raises the average feedback cost. This increase reflects improved coverage rather than reduced efficiency. With very small limits, feedback is concise but applies only to requests close to an \textit{allow} state, while larger limits improve feasibility at the expense of more complex and less interpretable feedback. The selected parameter values therefore balance coverage with feedback simplicity, as summarized in Table~\ref{tab:grid-optima}.

\subsection{Visibility Constraints in Effect}
\label{vis_ex}

\begin{figure}[h]
    \centering
    \begin{subfigure}{\linewidth}
        \centering
        \includegraphics[width=0.9\linewidth]{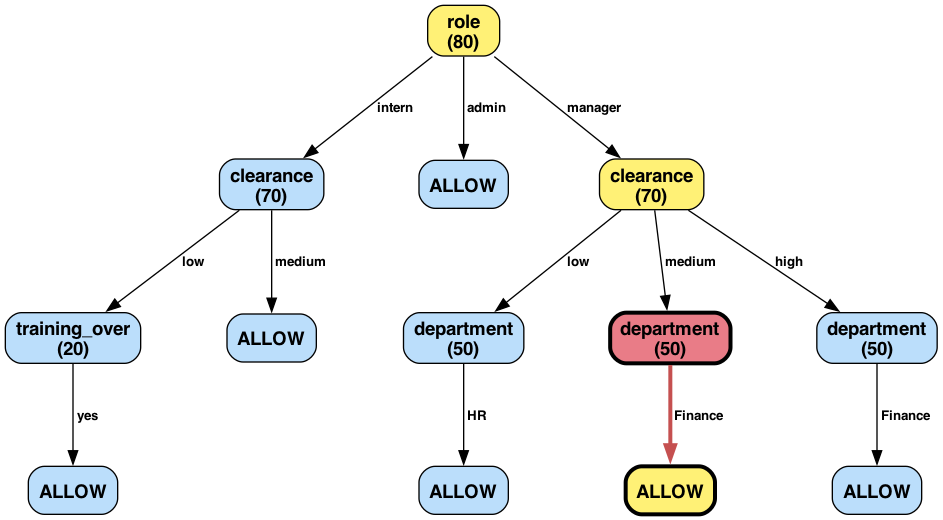}
        \caption{Without visibility constraints}
        \label{fig:no-vis}
    \end{subfigure}
    \hfill
    \begin{subfigure}{\linewidth}
        \centering
        \includegraphics[width=0.9\linewidth]{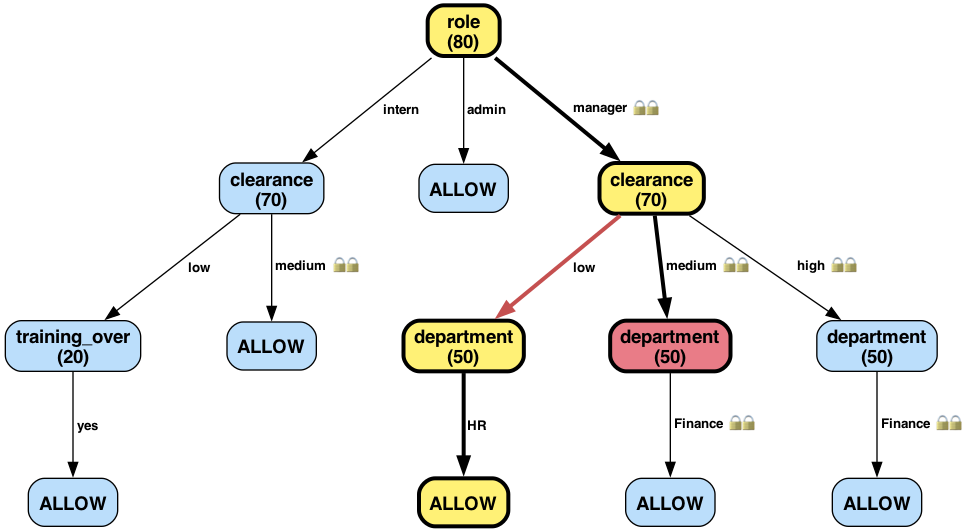}
        \caption{With visibility constraints}
        \label{fig:with-vis}
    \end{subfigure}
    \caption{Effect of visibility constraints.}
    \label{fig:visibility-example}
\end{figure}

To illustrate the effect of visibility constraints, Figure~\ref{fig:visibility-example} presents a concrete policy tree example. The request corresponds to a user with role=manager, clearance=medium, department=HR, and training\_over=yes, requesting access to a medium-sensitivity resource; the initial decision is \textsc{Deny}.
Without visibility constraints (Figure~\ref{fig:no-vis}), the minimum-cost explanation modifies the department attribute from HR to Finance, yielding a total cost of 50. When visibility constraints are applied (Figure~\ref{fig:with-vis}), specific attribute--value pairs (clearance=medium, clearance=high, and department=Finance) are hidden. Paths going through these predicates are now infeasible due to infinite traversal cost, eliminating the earlier optimal explanation. Feedback generation instead selects an alternative feasible path that modifies clearance from medium to low at a higher cost of 70.
This example shows how visibility constraints can be enforced in evaluation without changing policy structure.

\begin{table}[t]
\centering
\caption{Optimal parameters found through grid search.}
\label{tab:grid-optima}
\begin{tabular}{lccc}
\toprule
\textbf{Dataset} & \textbf{Optimal (max\_changes, max\_depth)}\\
\midrule
Synthetic-1 & (4, 30)\\
Synthetic-2 & (8, 30) \\
ABACLab Healthcare & (3, 5) \\
\bottomrule
\end{tabular}
\end{table}

\end{document}